\documentclass[a4paper,11pt]{article}
\usepackage{jcappub} 
\usepackage{amsmath,amssymb,amsthm,amsfonts}
\usepackage{orcidlink}
\usepackage{url}       
\usepackage{float}
\usepackage{xcolor}
\usepackage{graphicx}
\usepackage{wrapfig}
\usepackage{enumitem}
\usepackage[us,24hr]{datetime}
\usepackage{lineno}
\usepackage{color}
\usepackage{comment}
\usepackage{hyperref}
\usepackage{multicol}
\usepackage{titlesec}
\usepackage{fancyvrb}
\usepackage[nolist,nohyperlinks]{acronym}
\usepackage{csquotes}
\usepackage{lineno}
\usepackage{soul}
\usepackage{float}
\usepackage{booktabs}
\usepackage{multirow}
\usepackage{caption}
\usepackage{sidecap}
\usepackage{subcaption}
\usepackage{upgreek}
\usepackage[symbol,bottom]{footmisc}

\acrodef{UHE}{ultra-high-energy}
\acrodefplural{UHE}{ultra-high-energies}
\acrodef{UHECR}[UHECR]{ultra-high-energy cosmic ray}
\acrodefplural{UHECR}[UHECRs]{ultra-high-energy cosmic rays}
\acrodef{GMF}{Galactic Magnetic Field}
\acrodef{EGMF}{Extragalactic Magnetic Field}
\acrodef{EB}{Editorial Board}
\acrodefplural{EB}{Editorial Boards}
\acrodef{EAS}{extensive air shower}
\acrodefplural{EAS}{extensive air showers}
\acrodef{SGP}{supergalactic plane}
\acrodef{FD}{Fluorescence Detector}
\acrodefplural{FD}{Fluorescence Detectors}
\acrodef{SD}{Surface Detector}
\acrodefplural{SD}{Surface Detectors}
\acrodef{SBG}{starburst galaxies}
\acrodef{AGN}{active galactic nuclei}
\acrodef{DNN}{Deep Neural Network} 
\acrodefplural{DNN}{Deep Neural Networks}
\acrodef{1DCNN}[1-D CNN]{one dimensional Convolutional Neural Network} 
\acrodefplural{1DCNN}[1-D CNNs]{one dimensional Convolutional Neural Networks} 
\acrodef{CNN}[CNN]{Convolutional Neural Network} 
\acrodefplural{CNN}[CNNs]{Convolutional Neural Networks} 
\acrodef{LSTM}{Long Short-Term Memory}
\acrodefplural{LSTM}[LSTMs]{Long Short-Term Memory networks}
\acrodef{HPC}{High Performance Computing Center}
\acrodef{MURF}{Mines Undergraduate Research Fellowship}
\acrodef{ML}{Machine Learning}
\acrodef{HIM}{Hadronic Interaction Model}
\acrodefplural{HIM}[HIMs]{Hadronic Interaction Models}
\acrodef{GFN}[NFN]{Noise Flexible Network}
\acrodef{MSE}[MSE]{Mean Squared Error}
\acrodef{POEMMA}[POEMMA]{the Probe of Extreme Multi-Messenger Astrophysics}
\acrodef{PBR}[PRB]{POEMMA Balloon with Radio}
\acrodef{MEUSO}[M-EUSO]{the Multi-messenger Extreme Universe Space Observatory}
\acrodef{GCOS}[GCOS]{Global Cosmic Ray Observatory}

\newcommand{\lna}{$\ln{A}$}
\newcommand{\xmax}{$X_{\text{max}}$}

\newcommand{\Nmu}{$N_{\upmu}$}

\newcommand{\gcm}{g/cm$^2$}

\DeclareRobustCommand{\uvec}[1]{{%
  \ifcsname uvec#1\endcsname
     \csname uvec#1\endcsname
   \else
    \bm{\hat{\mathbf{#1}}}%
   \fi
}}

\title{Prospects for Deep-Learning-Based Mass Reconstruction of Ultra-High-Energy Cosmic Rays using Simulated Air-Shower Profiles}

\author[a,b]{Z.~Wang\,\orcidlink{0009-0008-5583-1887},}
\author[b,*]{E.~Mayotte\,\orcidlink{0000-0003-2618-9166}%
\note{Corresponding Author},}
\author[b]{S.~Mayotte\,\orcidlink{0000-0001-8957-8033},}
\author[a,b]{N.~Woo\,\orcidlink{0009-0006-4869-6264},}
\author[b]{J.~Burton-Heibges\,\orcidlink{0009-0001-4000-6576},}
\author[b]{N.~San Martin\,\orcidlink{0009-0007-6058-9549},}
\author[b]{and C.~Smith\,\orcidlink{0009-0003-5551-3153}}

\affiliation[a]{Colorado School of Mines, Department of Computer Science, Golden, Colorado}
\affiliation[b]{Colorado School of Mines, Department of Physics, Golden, Colorado}

\emailAdd{emayotte@mines.edu}

\abstract{
    Knowledge of the mass composition of ultra-high-energy cosmic rays is crucial to understanding their origins; however, current approaches have limited event-by-event resolution.
    With fluorescence telescope measurements of the longitudinal shower profile, there are opportunities to improve this situation by applying Machine Learning (ML) to leverage more information beyond \xmax{} alone.
    To our knowledge, we present the first study of a deep-learning neural-network approach to predict a primary's mass (\lna{}) directly from the longitudinal energy-deposit profile of simulated extensive air showers.
    We train and validate our model on simulated showers, generated with CONEX and EPOS-LHC, covering nuclei from $A = 1$ to $61$, sampled uniformly in \lna{}.
    After rescaling, our network achieves a maximum bias better than 0.4 in \lna{} with a resolution between 1.5 for protons and 1 for iron, corresponding to a proton-iron Merit Factor of 2.19 (AUC $= 0.976$).
    We benchmark this against simpler ML models trained on profile-shape parameters (\xmax{}, $E_{\rm cal}$, $R$, and $L$) extracted from the same data.
    We find that even simple models can substantially exceed published benchmarks for combinations of these observables, demonstrating that ML methods applied even to standard profile-shape parameters can significantly improve available mass sensitivity.
    The CNN outperforms this strong baseline, and
    this performance is only mildly degraded when cross-predicting on simulations made with the Sibyll-2.3d hadronic interaction model, showing robustness against model choice. The network also maintains its performance across a wide range of noise conditions.
    An ablation study further demonstrates that the full profile contains composition-sensitive structure not captured by the GH parameterization, while the strong performance of the CNN suggests this information should be resolvable in real events.}

\begin{document}
\renewcommand{\tableautorefname}{Tab.} 
\renewcommand{\sectionautorefname}{Sec.} 
\renewcommand{\subsectionautorefname}{Sec.} 
\renewcommand{\figureautorefname}{Fig.} 

\maketitle

\flushbottom
\section{Introduction}
\label{sec:introduction}
\Acp{UHECR}, cosmic rays with energies above 1\,EeV, are predominantly ionized atomic nuclei typically ranging in mass from protons at the lightest to iron at the heaviest~\cite{PierreAuger:2014sui}.
The measurement of the mass number, $A$, of \acp{UHECR}, and through it, their approximate charge, $Z$, is critical to their study.
Through rigidity, $R\propto E/Z$, the charge of a primary determines the degree to which cosmic magnetic fields deflect their trajectories, and strongly influences the maximum energy to which sources can accelerate.
Their mass, on the other hand, affects both the Lorentz factor and binding energy of primaries, and therefore governs the degree to which \acp{UHECR} are attenuated by background fields.
Together, these factors mean that the goal of identifying the sources of \acp{UHECR} relies on accurate knowledge of $A$ and $Z$.
Finally, to accurately probe fundamental physics using particle interactions of \acp{UHECR} in the atmosphere, knowledge of $A$ and the energy of nucleons participating in the initial interactions is essential.

Because \acp{UHECR} are extremely rare and are destroyed upon reaching the atmosphere, they are studied indirectly by observing the \ac{EAS} they cause~\cite{PierreAuger:2020kuy}. 
This means a mass measurement relies on inverting mass-sensitive signatures gathered as an \ac{EAS} moves through the atmosphere or when it strikes the ground.
The interpretation of these signatures is further complicated by two factors: the hadronic interaction models (e.g.,~\cite{PhysRevC.92.034906} and~\cite{PhysRevD.102.063002}) on which they rely carry large uncertainties, and many signatures have a meaningful random component, blurring any reconstruction of \ac{UHECR} mass.
These challenges mean no method has yet achieved an accuracy high enough to reconstruct mass on an event-by-event basis with a resolution sufficient for most single event studies~\cite{Coleman:2022abf}.

The evolution of the number of particles as a function of atmospheric depth (typically expressed as $X$ in units of \gcm{}) is referred to as the longitudinal profile of shower development.
Currently, the mass composition of cosmic rays is mainly estimated by picking out the atmospheric depth at which the shower reaches its maximum, \xmax{}~\cite{PierreAuger:2014sui}.
The number of muons in the shower when it reaches the ground, \Nmu{}, is also often used as it is highly correlated with primary mass~\cite{Kampert:2012mx}.
Neither \xmax{} nor \Nmu{} alone sufficiently constrains the primary particle type to allow for reliable identification on an event-by-event basis. 
Combining independent measurements of \xmax{} and \Nmu{} does, however, provide a means of enhancing the separation between lighter and heavier primaries, approaching the resolution needed for advanced studies~\cite{Coleman:2022abf}.
However, this combined approach is only realizable with hybrid detectors featuring large surface arrays, making it out of reach for proposed space-based observatories (e.g., \cite{POEMMA:2020ykm}), which can only make optical measurements of shower development in the atmosphere. 
To increase the resolving power of these optical measurements by fluorescence telescopes, this paper proposes a new approach, following in the early foot steps of studies like~\cite{Ambrosio:2005bs, Riggi:2007zz} and the recommendations of the UHECR whitepaper~\cite{Coleman:2022abf}, that uses machine learning to directly extract the mass of a primary particle from the longitudinal profile of an \ac{EAS}.
This approach enables all composition information in a profile to be used simultaneously.

\subsection*{Our Approach}
\label{sec:Approach}
\ac{EAS} shower profiles carry more information about composition than \xmax{} alone can provide.
Other variables, such as profile asymmetry and width, have already been shown to have composition sensitivity~\cite{Andringa:2011zz} and parameterized using data~\cite{PierreAuger:2018gfc}.
Combining these mass-sensitive variables into a single mass measurement has proven to be a challenging task.
Furthermore, there is no guarantee that these quantities fully encapsulate the mass sensitivity available in the shower profile.
\ac{ML} algorithms, however, have shown themselves to be adept at detecting subtle patterns in cosmic ray signals that have proven difficult for humans to observe~\cite{PierreAuger:2024flk}.
Therefore, we apply ML algorithms to simulated EAS profiles to test how much mass information can be extracted from them directly, and compare it to the performance achievable using known profile-shape parameters.

To perform this study, we generated a library of simulated \ac{UHECR} profiles with primaries ranging in mass from $A=1$ to $A=61$ using CONEX~\cite{Bergmann:2006yz, Pierog:2004re}.
From each event in this library, we extract three parameters to model the shower profile data collected by a fluorescence telescope in a real event: the atmospheric depth, $X$, in steps of 10\,g/cm$^2$, the energy deposition rate, $dE/dX$, at each step, and the zenith angle, $\theta$, of the event.
We then train \ac{ML} models to predict the natural log of mass of the primary cosmic ray \lna{}, using the above parameters as inputs.
To replicate a range of possible measurement conditions, we add two types of noise of varying strengths to degrade the measurements and limit the range of depths available in the shower profiles for the model to predict.
To validate our study, we will evaluate the \ac{ML} model using the noise-free original simulated data along with datasets featuring noise levels that are: lower than (to observe changes in performance), equivalent to (to reflect typical conditions), and higher than (to stress test the model), those typically encountered in real-world measurement. 
To further benchmark the \ac{ML} model, we will evaluate it against the `typical' noise conditions and compare its performance to that of linear discriminant~\cite{pedregosa2011scikit} and XGBoost~\cite{chen2016xgboost} models trained on the Gaisser-Hillas profile-shape parameters \xmax{}, $E_{\rm cal}$, $R$, and $L$ extracted from the same data.
We additionally perform an ablation study to quantify the composition-sensitive information present in the profile beyond what these parameters capture.

\section{Training Dataset}
\label{sec:dataset}
The cosmic ray shower profiles used to train the machine learning model are generated with CONEX using the hadronic interaction models EPOS-LHC and Sibyll 2.3d~\cite{PhysRevC.92.034906, PhysRevD.102.063002}. 
We generate these shower profiles by sampling primaries from a uniform distribution in \lna{} rather than $A$.
We chose to sample flatly in \lna{} primarily for two reasons. First, because most mass-sensitive shower observables, such as \xmax{}, \Nmu{}, and even shower footprint timing parameters~\cite{PierreAuger:2017tlx}, scale well with \lna{}. 
Second, a logarithmic sampling allocates more events to the light‑mass region (e.g., proton vs. helium) where small differences have high impact without wasting statistics or network attention on the nearly indistinguishable heavy nuclei (e.g., manganese vs.\ iron). 

Specifically, we draw a continuous value
\begin{align}
  \ln A_{\text{rand}}
  = \ln A_{\min}
    + U\!\bigl(0,\,\ln A_{\max}-\ln A_{\min}\bigr),
\end{align}
where $U(0,x)$ returns a random number uniformly distributed between $0$ and $x$, and $A \in [1,61]$ for EPOS-LHC (5 greater than iron to avoid edge effects). For Sibyll, $A \in [1,56]$ due to the model being unable to simulate masses greater than iron ($A=56$).
The mass number $A_{\mathrm{rand}}$ is then calculated as
\begin{align}
    A_{\mathrm{rand}} = \exp{\bigl(\ln A_{\mathrm{rand}}\bigr)}\,.
\end{align}
and is rounded to the nearest integer.
Lastly $A_{\text{rand}}$ is converted to a CONEX particle code:
\begin{align}
 p_{\text{CONEX}}= A_{\text{rand}} \cdot 100\,.
\end{align}

\begin{figure}[H]
    \vspace{-2mm}
    \centering
    \includegraphics[width=\textwidth]{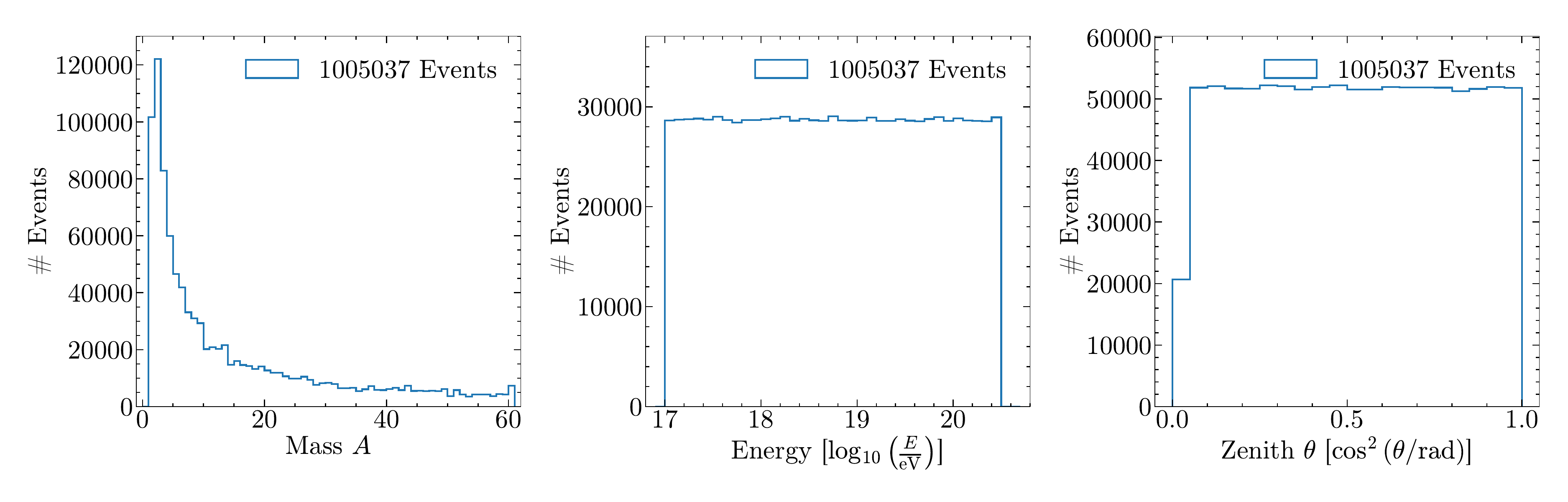}
    \vspace{-10mm}
    \caption[]{CONEX EPOS-LHC simulations.
    Left: The mass distribution of simulated primaries, generated to be flat in \lna{} ($A\in [1,61]$). 
    Middle: The flat in $E^{-1}$ energy distribution ($E \in [10^{17},10^{20.5}]$\,eV). 
    Right: The flat in $\cos^2{\theta}$ zenith distribution ($\theta \in (0^\circ,80^\circ)$).
    }
    \label{fig:FlatMCDistributions}
    \vspace{-2mm}
\end{figure}

Using a shower library constructed in this way allows us to train a network to perform the regression task of predicting a continuous \lna{} value, rather than the classification task of sorting events into a discrete mass group, which may not accurately represent the UHECR flux.
This library, therefore, avoids potential bias from predefined composition priors and yields a direct, uncertainty-preserving reconstruction of the primary mass.
Both simulation sets also cover a zenith angle range $\theta \in (0^\circ,80^\circ)$ that is flat in $\cos^2{\theta}$, with an energy distribution $E \in [10^{17},10^{20.5}]$\,eV that is flat in $E^{-1}$.
The distributions of the key generation variables for the EPOS-LHC library are shown in \autoref{fig:FlatMCDistributions}.

We produced two libraries for this study, each containing one million simulated CONEX showers, one generated with EPOS-LHC and the other with Sibyll.
The values for $X$, $dE/dX$, $\theta$, and $A$ from each shower in these libraries were then extracted and used as the input data for the networks described later. 
Because EPOS-LHC allowed for a broader range of masses, it was used as the model on which the network was trained and validated, while Sibyll was used only in predictions to qualify hadronic interaction model dependence.

\section{Network Architecture}
\label{sec:networkarchitecture}
Though the mass composition of \acp{UHECR} consists of distinct, discrete values, there are compelling advantages to framing it as a regression task rather than a classification problem. 
First, the mass of \ac{UHECR} primaries exhibits an inherent ordinal relationship. 
Misclassifying a heavy nucleus (e.g., iron) as a light one (e.g., proton) is a more significant error than confusing two light nuclei (e.g., proton and helium).
Although classification techniques, such as ordinal categorical cross-entropy loss, exist to account for such hierarchy, they ultimately rely on distance-based weighting schemes that mirror the assumptions of a continuous mass spectrum. 
Second, a regression model naturally supports this continuous interpretation, while allowing for the output to be discretized via rounding, flooring, or ceiling in classification as desired. 
This approach also offers greater resolution as regression enables finer distinctions when two predicted masses are close.
It also allows for a more interpretable model, particularly in the context of systematic trends or explainability studies. 
We adopt a regression framework over a purely categorical classification for these reasons.

Given the temporal nature of the task, we explored several architectures commonly used for time-series data.
These included \ac{LSTM} networks, \ac{1DCNN}, and Transformer models. 
\acp{LSTM} are well-suited to capturing long-range temporal dependencies, which may be relevant for understanding how energy deposition across the shower profile relates to primary mass. 
\acp{1DCNN} treat the shower profile as a sequential spatial signal, offering computational efficiency through parallelization. 
Transformer models can successfully model time-series tasks and global dependencies via attention mechanisms.
Additionally, to establish a baseline on combinations of known shower parameters we also used a linear discriminant method chosen for it simplicity and XGBoost chosen as a moderately complex and flexible model.

This study focuses primarily on the \acsu{CNN} architecture, as alternative models—LSTM and Transformer networks—either underperformed or required substantially greater computational resources. 
We made this selection based on performance on noise-free data, assuming that these results represent an upper bound on achievable model accuracy. 
Empirically, the LSTM model exhibited poor generalization and stagnated during extended training.
The Transformer achieved comparable accuracy to the \ac{CNN}, but at a significantly higher computational cost. 
The Transformer also struggled to train when noise was applied to the datasets. 

\autoref{tab:CNN_architecture} outlines the structure of the \ac{CNN} employed for the mass reconstruction task.
This pattern of the \ac{CNN}, increasing filter counts while reducing the resolution, is commonly seen in many \ac{CNN} architectures such as VGG \cite{Simonyan:2014cmh} and ResNet \cite{He:2015wrn}. 
The underlying concept is to progressively develop a rich and abstract representation of the data by initially capturing local features and then integrating them into increasingly complex patterns as the network deepens. 
In the initial blocks, the network uses a small number of filters (e.g., two filters) and the goal is to capture fine-grained, local features from the input (such as small-scale variations in the signal).
Since features at this level are relatively simple, a smaller number of filters is sufficient, and keeping the representation low-dimensional helps to control the complexity and computation required at the early stage. 
As the signal passes through successive layers with pooling, reducing its temporal resolution, the receptive field of the network effectively increases. 
Increasing the number of filters (e.g., from 2 to 4, then 8, 16, 32, and finally 64) enables the network to encode a richer and more diverse set of feature maps, which are crucial for capturing the complexity in \ac{UHECR} reconstruction. 
A pooling layer is used after each block of convolutional layers to reduce the temporal dimension, lowering the computational costs in deeper layers. 
Finally, a flatten layer collapses extracted features into a one-dimensional vector, while the dense layer maps these features to the mass prediction. 

\begin{table}[!htbp]
\centering
\begin{tabular}{ll|cccc}
Block & Layer (Type) & Kernel/Pool Size & Filters & Activation & Output Shape \\
\midrule 
\midrule 
\multirow{2}{*}{Block 1} & Conv1D (7 layers) & 2 & 2  & ELU    & (None, 700, 2)\\
                         & MaxPooling1D      & 2 & -- & --     & (None, 350, 2)\\
\midrule
\multirow{2}{*}{Block 2} & Conv1D (6 layers) & 2 & 4  & ELU    & (None, 350, 4)\\
                         & MaxPooling1D      & 2 & -- & --     & (None, 175, 4)\\
\midrule
\multirow{2}{*}{Block 3} & Conv1D (5 layers) & 2 & 8  & ELU    & (None, 175, 8)\\
                         & MaxPooling1D      & 2 & -- & --     & (None,  87, 8)\\
\midrule
\multirow{2}{*}{Block 4} & Conv1D (3 layers) & 2 & 16 & ELU    & (None,  87,16)\\
                         & MaxPooling1D      & 2 & -- & --     & (None,  43,16)\\
\midrule
\multirow{2}{*}{Block 5} & Conv1D (2 layers) & 2 & 32 & ELU    & (None,  43,32)\\
                         & MaxPooling1D      & 2 & -- & --     & (None,  21,32)\\
\midrule
Block 6                  & Conv1D (1 layer)  & 2 & 64 & ELU    & (None,  21,64)\\
\midrule
                         & Flatten           & --& -- & --     & (None, 1344)  \\
                         & Dense             & --& -- & Linear & (None,    1)  \\
\end{tabular}
\caption{Architecture of the implemented \ac{CNN}}
\label{tab:CNN_architecture}
\end{table}

For various reasons, kernel sizes are deliberately kept small in the \ac{CNN}. 
First, it forces each convolutional layer to initially focus on fine-grained patterns, encouraging the model to learn simple features. 
As the network deepens, larger kernels are unnecessary because the pooling layers aggregate local features into progressively abstract representations by reducing the temporal dimension. 
Second, smaller kernels reduce the number of trainable parameters, lowering the computational cost of training and the risk of overfitting. 
For the convolutional layers, we chose ELU as the activation function, as we observed that ELU outperforms ReLU and some other ReLU-like activations in tests. 
ELU delivered superior performance likely because it helps mitigate the vanishing gradient problem, accelerates training, and leads to better generalization compared to other ReLU-like activation functions \cite{clevert2016fastaccuratedeepnetwork}.
This network architecture, resulting from extensive iterative tuning, shows promising performance for the mass reconstruction task.
However, other architectures or further optimizations would likely result in higher performance than the above-described \ac{CNN} can produce.

\section{Pre-processing and Noise Simulation}
\label{sec:prep}
We pre-process the data to fit it to the network architecture described above and to improve training performance by reducing the size of the numbers used as network inputs and increasing feature accessibility.
First, because the length of a CONEX profile changes with zenith angle (due to a 10\,g/cm$^2$ depth step-length), we standardize the length of the $X$ and $dE/dX$ input arrays.
The zenith angle range of generated events was 0$^\circ$ to 80$^\circ$, resulting in  $\sim$200 to 700, 10\,g/cm$^2$, steps per file.
We therefore zero-pad the end of $X$ and $dE/dX$ arrays until they contain exactly 700 values corresponding to a maximum profile length of 7000\,g/cm$^2$.

Next, we reduce the magnitude of the numbers in the $dE/dX$ data to increase efficiency and feature accessibility. 
Two options were considered: a linear rescale by expressing the energies in GeV, or expressing values in units of $\log_{10}(E/\rm{eV})$.
In training, the linear scaling underperformed, so the log transformation was selected. 
This may be because the log-transformed values, on average, exhibit higher variance under min-max scaling, both in clean profiles (0.0033 linear vs. 0.088 log) and in profiles with typical noise (0.0031 linear vs. 0.11 log), which theoretically should transfer to a higher rate of learning and a higher degree of feature recognition. 
Liner scaling is, however, used in \autoref{fig:noise_comparison} as we find it provides for a superior visualization. 

\subsection*{Noise Models}
The simulated shower profiles are modified in two ways to model conditions seen in measurement:
A \emph{baseline} is added to blind the network to the first interaction point, model a constrained field of view, and to mimic atmospheric background light, and \emph{Gaussian noise} to model the statistical fluctuations in light production, propagation, and detector response.

\paragraph*{Baseline}
The baseline is set as a percentage of the profile's peak $dE/dX$ value. 
When it is applied to the simulated profiles, each $dE/dX$ value is replaced by the larger of the original value and the baseline, rather than adding the baseline.
This choice allows the baseline to simulate both a constrained field of view and the presence of atmospheric light.
An additive baseline uniformly shifts all $dE/dX$ values upward without fully suppressing the information in low-signal regions, thereby potentially preserving information that would be inaccessible in a constrained measurement scenario. 
Accordingly, we saw that applying an additive baseline produced higher mass separation, but we viewed this improvement to be artificial as the network’s attention to the low-signal region was increased relative to the chosen approach, which indicated that additive baselines preserved otherwise inaccessible information.

The baseline height varied from 5\,\% to 75\,\% of the peak $dE/dX$ value.
This direct setting of the baseline as a fraction of the peak $dE/dX$ allows for an efficient test of a range of possible signal-to-noise ratios at all simulated energies. 
This has the added benefit of mostly removing energy as a parameter of performance, allowing us to reduce the parameter space in which we need to qualify performance to only noise levels and primary species.
This baseline served a secondary purpose of being an effective means of trimming the profiles to reduce the depth ranges observed while keeping \xmax{} within the field of view, though \xmax{} is always well contained, which does not reflect the conditions always encountered in measurement.

\paragraph*{Gaussian Noise}
The Gaussian noise is generated by sampling a Gaussian distribution with a mean of 0 and a standard deviation set as a percentage of the peak $dE/dX$ value at all 700 steps in $X$. 
The resultant values are then summed with the values in the simulated profile and baseline.
We benchmark on Gaussian noise levels ranging from $2.5\,\%$ to $20\,\%$ of peak. 
This approach effectively simulates the variations and uncertainties observed in actual fluorescence telescope measurements, while again allowing for a highly efficient generation of a wide range of possible noise levels.

\begin{figure}[H]
    \centering
    \begin{subfigure}[b]{0.48\textwidth}
        \centering
        \includegraphics[width=\textwidth]{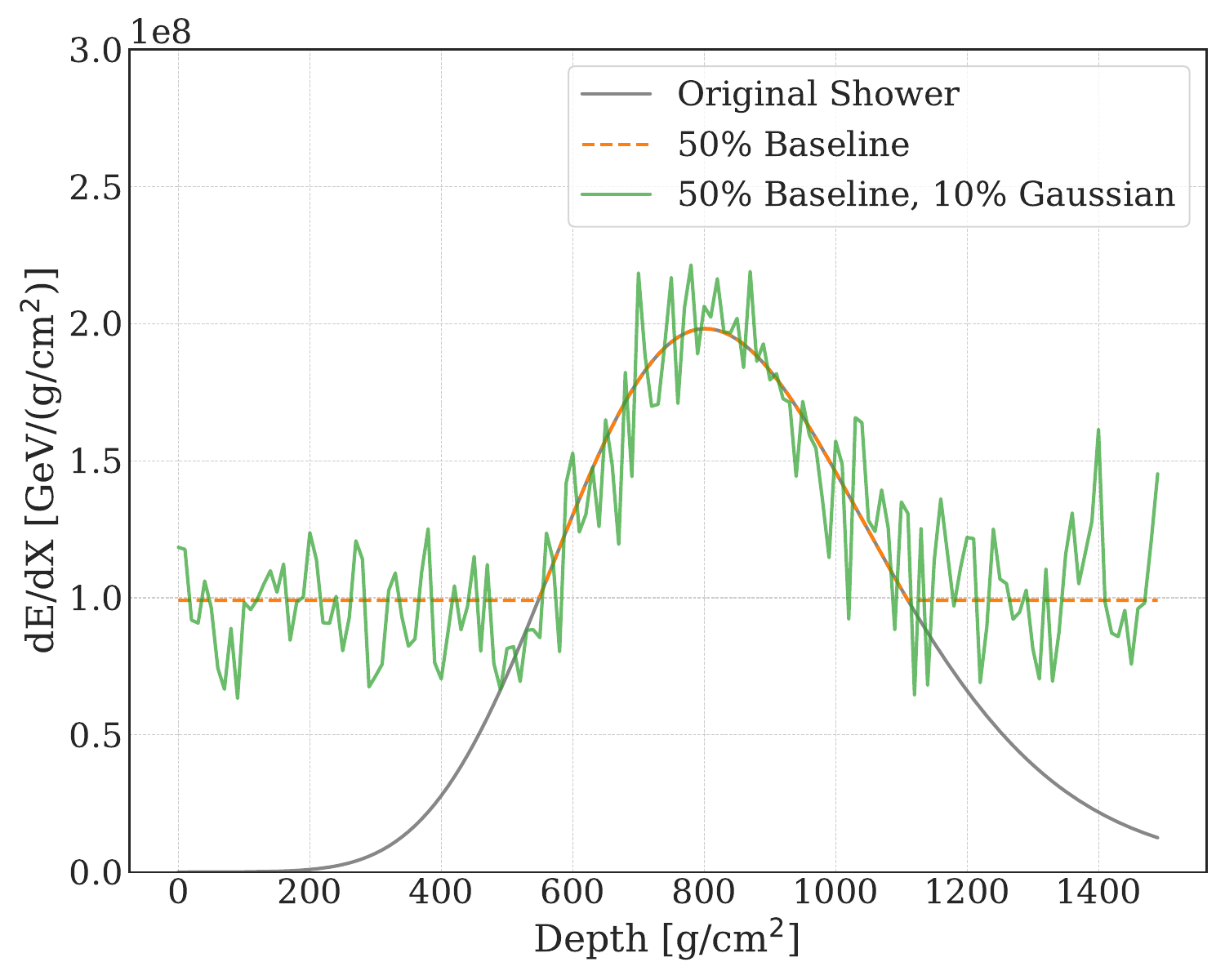}
        \caption{Noise level taken as typical in observations.}
        \label{fig:auger_noise_profile}
    \end{subfigure}
    \hfill
    \begin{subfigure}[b]{0.48\textwidth}
        \centering
        \includegraphics[width=\textwidth]{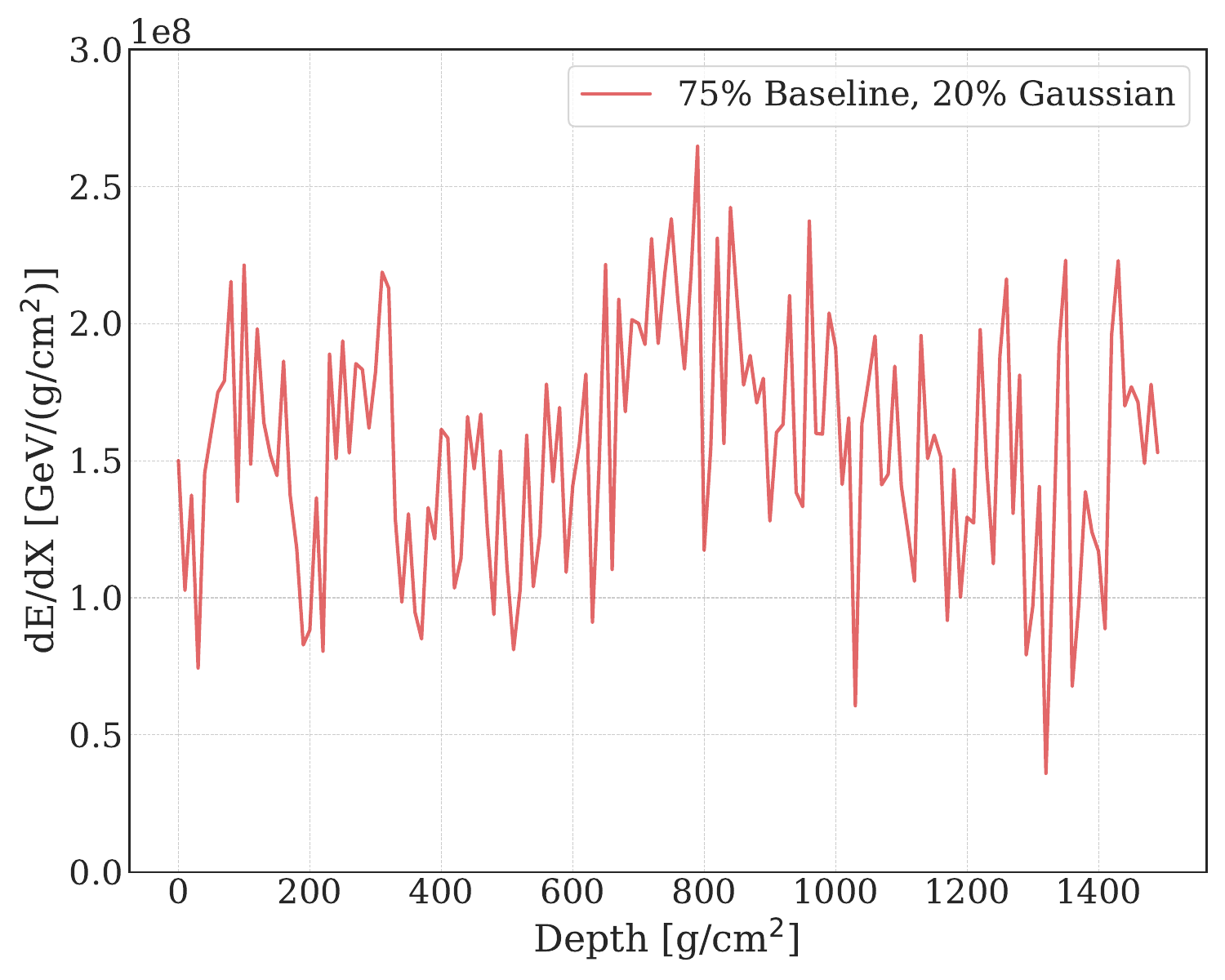}
        \caption{Highest noise level evaluated.}
        \label{fig:highest_noise_profile}
    \end{subfigure}
    \caption{Simulated profiles under different evaluated noise conditions.}
    \label{fig:noise_comparison}
\end{figure}

By combining the above-defined baseline and Gaussian noise types, shower profiles can be generated that closely resemble those observed in actual measurements of the shower profile, such as those available in the open data of the Pierre Auger Observatory \cite{auger_open_data_v3_2024}.
An example of a typical-looking shower profile with a $50\,\%$ baseline and $10\,\%$ Gaussian noise can be seen in \autoref{fig:auger_noise_profile}. 
For the remainder of the paper, this combination will be referred to as \emph{typical noise}. 
For initial training and comparison of the networks, a $50\,\%$ baseline and $5\,\%$ Gaussian noise was used.
This moderate level of baseline and Gaussian noise was chosen to strike a balance between the clarity of mass-sensitive features in a profile and the network's ability to generalize against higher noise conditions.
\autoref{fig:highest_noise_profile} shows an example where the highest level of both the baseline and the Gaussian noise is applied. 
In all cases, noise values that would reduce $dE/dX$ below zero were clipped to $1\,\mathrm{keV/(g\,cm^{-2})}$.
This lower bound corresponds to the minimum energy observed in our libraries beyond the first point of interaction and ensures compatibility with the logarithmic scaling of $dE/dX$ used in the study.

\section{Network Training and Performance}
\label{sec:Base-training}

The 1 million events of the EPOS-LHC data set were split 70\,\%/10\,\%/20\,\% for training, validation, and testing, respectively, while the entirety of the Sibyll set was reserved for testing \ac{HIM} dependence.
This partitioning of the data left 700,000 events with which to train the networks to extract \lna{} from the noisy simulated shower profiles.
Training was carried out on a dedicated workstation with two NVIDIA A6000 ADA GPUs and enough RAM to hold the whole training dataset in memory. 
With this setup, the \ac{CNN} was trained at a rate of roughly one epoch per minute.

This high training speed allowed for performing long training sessions on a variety of architectures, data preparations, and noise levels. 
Generally, training completion was determined when performance plateaued for 100 epochs or when a significant increase in training loss was observed. 
The model that yielded the lowest overall validation loss during training was selected as the final model.
Over the many long training runs of this study, a limit of 2500 epochs appeared to be a safe upper boundary, as the majority of networks satisfied one of the above completion criteria well before this point.

\subsection{Gauging Performance}
\label{sec:baseperformance}

Metrics are required to evaluate the performance of a model and compare it to other published methods.
In this paper, a few quantities are used.
Because we aim to gauge the potential for directly predicting \lna{} from an observed shower profile, we will present the mass resolution, $\sigma(\Delta \ln A)$, and mean prediction bias, $\langle \Delta \ln A \rangle$ of our final model's predictions for all masses between $A=1$ and $A=61$.
However, as will be shown in \autoref{sec:NFNAccuracy}, unless corrected, networks can artificially squeeze their predictions to improve precision at the cost of accuracy.
This means it is useful to have a representative metric that is less sensitive to this effect because it compares the distance between the species’ mean values to the combined width of their distributions, and can therefore be used to more robustly gauge each tested network's ability to distinguish between different mass species.
For this, we choose the \emph{Figure Of Merit (FOM)} which is defined as:
\begin{align}
    FOM(A_1,A_2) = \frac{\left| \mu_{A_1} - \mu_{A_2} \right| }{\sqrt{\sigma_{A_1}^2 + \sigma_{A_2}^2}}\,,
\end{align}\label{eq:mf}%
where $\mu_{A_1}, \mu_{A_2}$ and $\sigma_{A_1}, \sigma_{A_2}$ are the mean and standard deviation of the distributions of the mass-sensitive quantity under study, with $A_1$ and $A_2$ denoting the mass number of the tested species.
The final step requires selecting two representative species $A_1$ and $A_2$ for comparison.
For these, we select the standard benchmarks of proton and iron.
The proton--iron Figure Of Merit, $FOM(1,56)$, is commonly used as a benchmark of mass sensitivity in the literature (e.g.,~\cite{Flaggs:2023exc}), and to quantify composition-separation performance for proposed next-generation observatories (see \cite{Coleman:2022abf}).
Therefore, it is useful for comparing the performance of our networks with other established methods.
In what follows, when quoting a single number for the Figure Of Merit, we focus on $FOM(1,56)$ and refer to it as the \emph{Merit Factor} unless stated otherwise.

To provide a context with which to understand the performance of the network, it is useful to compare the Merit Factors achieved in this study to what is possible using the commonly used mass-sensitive observables.
Of particular interest are the accessible mass-sensitive profile parameters \xmax{}, $R$, and $L$ (excluding the first point of interaction, which is not realistically observable).
At 10\,EeV using \xmax{} alone, a resolution of $20$\,g/cm$^2$ yields a Merit Factor of $1.41$ \cite{Flaggs:2023exc}, while perfect knowledge of \xmax{} has been shown to yield a Merit Factor of $1.5$ representing the maximum limit achievable using only \xmax{}~\cite{PierreAuger:2021xnt}.
Combining the commonly reconstructed profile parameters of \xmax{}, $R$, and $L$ at set energies at once with realistic experimental resolutions yields a Merit Factor of $1.51$, while their combination with perfect knowledge provides an upper-bound benchmark Merit Factor of $1.83$~\cite{Flaggs:privcomm}.
This provides an estimate of the level of mass separation achievable with traditional profile-based methods.
If we additionally look beyond profile parameters to those that can be measured by ground arrays, Merit Factors as high as 1.8 are achievable by leveraging \Nmu{} alone \cite{Flaggs:2023exc}.
An independent combination of \xmax{} and \Nmu{} can exceed a Merit Factor of 2, which therefore represents the planned mass reconstruction approach of current and next-generation ground-based observatories~\cite{Coleman:2022abf}.
From \cite{Flaggs:2023exc}, simultaneous exact knowledge of all currently used mass-sensitive parameters allows for Merit Factors above 2.5.
These published values provide useful external context; however, because they were obtained under different conditions and analysis frameworks, in \autoref{sec:noisefree} and \autoref{sec:GHComparison} we additionally perform direct internal comparisons using models trained on profile shape parameters extracted from the same simulated dataset.
This allows us to more robustly isolate the CNN's advantage over the profile-shape parameters themselves, independent of the method used to combine them.

As a complementary metric, we also report the area under the receiver operating characteristic (ROC) curve (AUC)~\cite{FAWCETT2006861}.
The ROC curve is constructed by treating a predicted quantity as a classification threshold and scanning it to obtain the true positive rate and false positive rate at each threshold.
For this paper, proton and iron events are selected from the test set using their MC truth mass number, and the predicted \lna{} value is scanned from low (proton-like) to high (iron-like) values, yielding at each point a proton true positive rate (TPR) and false positive rate (FPR).
TPR is then plotted as a function of FPR to produce the ROC curve and the AUC can be extracted.
The AUC provides a threshold-independent measure of separation power that, unlike the Merit Factor, is invariant under monotonic transformations of the predicted variable.
This property is important because the $FOM$ can be inflated by non-linear variable transformations when models are not constrained across the full mass spectrum; this artifact and its suppression in our study are discussed in \autoref{sec:NFNPerformance}, where AUC and Merit Factor are shown to rank all models identically.
Throughout this paper, both Merit Factor and AUC are reported together to provide a literature-comparable benchmark and a transformation-robust measure of mass discrimination power.

Finally, while proton and iron are generally chosen as they represent the lightest and heaviest elements expected to appear in the \ac{UHECR} flux, it is expected that events similar in energy will likely come from mass groups that are much closer together (proton and helium, for example).
For these reasons, in \autoref{sec:NFNPerformance} we also present the final resolution at which \lna{} can be predicted over the range of tested \lna{} values.
Additionally, in \autoref{tab:MFTable} and \autoref{fig:confusion_matrix}, the $FOM$ for intermediate mass comparisons is shown.

\subsection{Predicting on Noise-Free Data}\label{sec:noisefree}
\begin{figure}[!htb]
    \centering
    \includegraphics[width=0.9\linewidth]{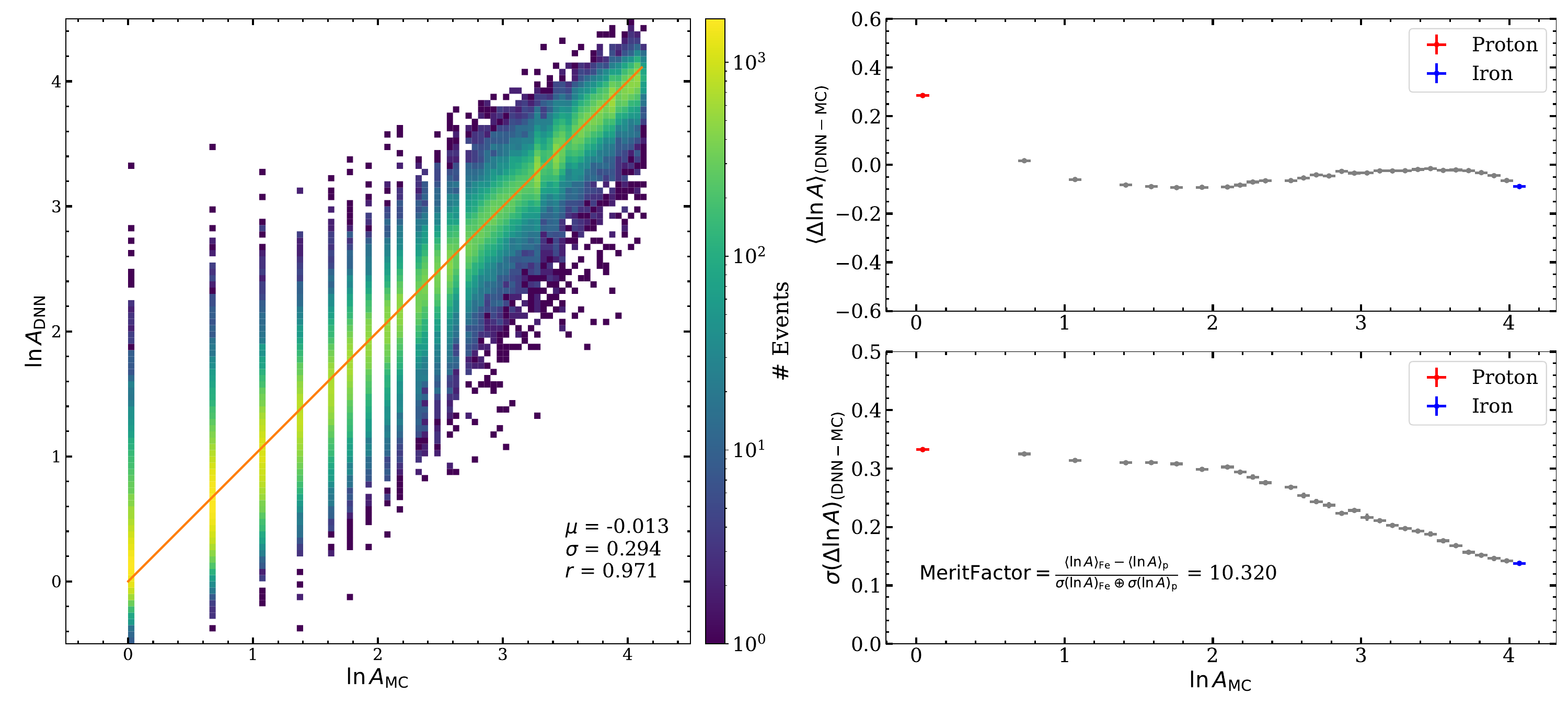}
    \caption{Network Performance on noise-free data}
    \label{fig:NoNoise}
\end{figure}

To establish an absolute upper bound on the achievable performance of our approach, the network was initially trained on noise-free data.
\autoref{fig:NoNoise} shows the peak performance achieved by the network under these idealized conditions, where very high sensitivity to composition was found, with the Merit Factor reaching a value of 10.32.
To understand this extreme result, we examined the trained model's attention and found that it weighted the information in the first $\sim 20$ bins far above the information elsewhere in the shower.
This led us to suspect that the first interaction was providing the bulk of this information, and we proceeded to blind the network to this information using baseline noise, as it would not be accessible in real measurements.

Out of curiosity, we also wanted to identify what piece of information was responsible for these high values.
One suggestion we received was that the rate of energy deposition prior to the first point of interaction would be proportional to the parent's charge and that the network could use this to determine the primary's mass with high fidelity.
To test this hypothesis, we removed the information before the first interaction ($X_{\rm first}$) by setting these bins to a uniform value of 1 \,eV/(g\,cm$^{-2}$).
This treatment reduced the Merit Factor to 6.8, indicating that early-depth features do play a significant role.
Extending this treatment to the first 20 bins resulted in a further reduction of the Merit factor to 5.2, after which the heavy weighting of model attention to the first bins was no longer observed.

To provide a direct baseline against which to compare the noise-free CNN performance, we also trained the linear discriminant and XGBoost models to predict $\ln A$ from the four Gaisser-Hillas (GH) parameters \{\xmax{}, $\, E_{\rm cal},\, L,\, R\}$.
The GH parameters were extracted from the same truncated noise-free profiles following the fitting procedure of~\cite{PierreAuger:2018gfc}, and are considered as close stand ins for MC truth values.
The calorimetric energy $E_{\rm cal}$ is used in place of the total shower energy as it is directly available from the longitudinal profile, preserves mass dependence, and allows us to avoid invisible-energy corrections.
The linear discriminant model was chosen as the simplest possible combination of variables, while XGBoost was selected to provide a strong non-linear baseline.
The results for all trained models are shown in \autoref{tab:NoiseFreeComparison}.
The CNN substantially outperforms both GH parameter models, establishing an upper bound on the improvement available from direct profile regression relative to extracted parameter regression.

\begin{table}[!htb]
\centering
\caption{Performance of models on MC truth data. The table compares CNN performance on clean data with the bins prior to $X_{\rm first}$ blinded, to the performance of Linear Discriminant and XGBoost on MC truth values of \xmax{}, $E_{\rm cal}$, $L$, and $R$.}
\label{tab:NoiseFreeComparison}
\renewcommand{\arraystretch}{1.3}
\begin{tabular}{lcc}
\hline\hline
Model & Merit Factor & AUC \\
\hline
Linear Discriminant (4 par.) & 1.69 & 0.983 \\
XGBoost (4 par.)             & 2.83 & 0.992 \\
CNN (full profile)                                        & 6.83 & 0.9999 \\
\hline\hline
\end{tabular}
\end{table}

Notably, even the XGBoost model operating on only four extracted parameters achieves a Merit Factor of 2.83, already exceeding the published benchmarks for any single observable and for the combination of \xmax{}, $R$, $L$ from~\cite{Flaggs:2023exc}, which implicitly included energy information due to it being calculated individually for each energy bin.
This suggests that applying modern ML methods even to standard profile-shape parameters can yield substantial gains in mass sensitivity, providing an immediately viable avenue to mass sensitivity improvement in current observatories.
With these CNN results established as the noise-free upper bound, we proceeded to train networks on data modified with our noise model. The result of that work is the subject of the rest of the paper.

\subsection{Benchmarking Network Prospects for Different Noise Conditions}
\label{sec:baseperformnacevsNoise}
Our general approach for developing a model that can successfully predict on noisy data was to first train a base model on a high 50\,\% baseline and mild 5\,\% Gaussian noise level to tune network parameters and data preparation. 
This trained base model then served as the foundation from which new networks could be trained via transfer learning. 
Once fully trained, the base model achieved a Merit Factor of 2.3.

After establishing the base model, the next step was to benchmark the performance of the method across a range of noise levels. 
To this end, we proceeded to train and benchmark separate networks on combinations of baseline and Gaussian noise.
For this, five baseline levels (5\,\%, 10\,\%, 25\,\%, 50\,\%, and 75\,\% of peak) and four Gaussian noise levels (2.5\,\%, 5\,\%, 10\,\%, and 20\,\% of peak) where used resulting in a total of 20 benchmarked noise conditions.

To train these networks, transfer learning was employed, as networks trained from scratch exhibited significantly slower convergence and were occasionally terminated early due to training breakdowns.
Transfer learning from the base model enabled faster and more stable convergence on the new noise condition, allowing the network to focus primarily on fine-tuning and capturing the nuances specific to each noise condition.
Each of the models was trained for an additional 1500 epochs on top of the base model.
Again, the model with the lowest validation loss is picked as the best-performing model for that noise level.

It is important to note that because the baseline height and Gaussian noise level are defined as a fraction of a profile's peak $dE/dX$, a network could exploit this correlation to gain a precise estimate of the peak amplitude by looking at low $X$ values.
This, in turn, could be used to give the network a better handle on the energy of the shower, which could improve the prediction of \lna{}.
Because of this, the results in this section are intended only to characterize how the network architecture responds to controlled increases in noise-induced degradation under fixed observational constraints and should not be interpreted as final physics performance.
Later, in \autoref{sec:FlexibleModel}, the baseline and Gaussian noise levels are randomized for every event, explicitly removing this potential bias avenue.
This allows the results of that section to stand alone as a gauge of the possible physics performance of the method. 

\paragraph{Performance}
The performance of the network architecture against varying noise conditions is shown in \autoref{tab:EPOS_Best_Performance}.
The network achieves Merit Factors above 2.0 at all but the most severe noise conditions, with performance degrading gradually as noise increases.
Specifically, within each baseline level, the expected degradation of performance as the Gaussian noise level increases is observed.
A similarly expected trend of worsening performance is also seen when the Gaussian noise level is fixed while the baseline height is increased.
Interestingly, however, the rate of degradation is much slower for baseline height increase than it is for Gaussian noise increase, indicating that most of the key mass information being picked up is conveniently near the brighter part of the shower. 

\begin{table}[!htb]
    \centering
    \caption{Performance of the network on different \acsp{HIM} and noise conditions (Merit Factor).}
    \label{tab:Peak_Performance_Comparison}
    \begin{subtable}[b]{0.49\textwidth}
        \centering
        \caption{Performance on native EPOS-LHC data}
        \label{tab:EPOS_Best_Performance}
        \begin{tabular}{ccccc}
            \toprule
            \multirow{2}{*}{\shortstack{Baseline\\Height}} & \multicolumn{4}{c}{Gaussian Noise} \\
            \cmidrule(lr){2-5}
             & 2.5\% & 5.0\% & 10\% & 20\% \\
            \midrule
            5\%  & 2.57 & 2.39 & 2.19 & 2.00 \\
            10\% & 2.55 & 2.37 & 2.24 & 2.03 \\
            25\% & 2.41 & 2.34 & 2.16 & 1.97 \\
            50\% & 2.36 & 2.27 & 2.20 & 1.88 \\
            75\% & 2.29 & 2.23 & 2.05 & 1.57 \\
            \bottomrule
        \end{tabular}
    \end{subtable}
    \hfill
    \begin{subtable}[b]{0.49\textwidth}
        \centering
        \caption{Cross-validation on Sibyll data}
        \label{tab:Sibyll_Best_Performance}
        \begin{tabular}{ccccc}
            \toprule
            \multirow{2}{*}{\shortstack{Baseline\\Height}} & \multicolumn{4}{c}{Gaussian Noise} \\
            \cmidrule(lr){2-5}
             & 2.5\% & 5.0\% & 10\% & 20\% \\
            \midrule
            5\%  & 2.37 & 2.22 & 2.05 & 1.90 \\
            10\% & 2.31 & 2.19 & 2.05 & 1.89 \\
            25\% & 2.22 & 2.15 & 2.02 & 1.89 \\
            50\% & 2.15 & 2.10 & 2.01 & 1.83 \\
            75\% & 2.10 & 2.05 & 1.94 & 1.59 \\
            \bottomrule
        \end{tabular}
    \end{subtable}
\end{table}

\subsection{Estimating Hadronic Interaction Model Dependence}
\label{sec:BaseHIM}
The training of these networks is reliant on simulated showers generated using \acfp{HIM}.
\acp{HIM} rely on extrapolating physics data from the LHC to higher energies and, therefore, carry significant uncertainties in their results. 
Critically, because of this, the networks inherit the uncertainties of these models, and therefore their predictions are model-dependent. 
To investigate this model dependence, the above-described models trained on EPOS-LHC have been used to predict on the Sibyll dataset.

The performance of the EPOS-LHC trained models on Sibyll showers is shown in \autoref{tab:Sibyll_Best_Performance} for a range of noise conditions.
Surprisingly, there is only a slight decrease in the overall performance compared to the native EPOS-LHC predictions, and again, the network achieves Merit Factors above 2.0 at all but the most severe noise conditions.
However, the predictions for all masses higher than that of protons moved to lighter values, resulting in the lower overall Merit Factors.
These results validate the network architecture and suggest that the mass ordering and separation powers of the models are robust against \ac{HIM} uncertainties. 
These results, however, present a challenge to the method's ability to make a robust measurement of absolute primary mass.
Because there is, as of yet, no model-independent direct measurement of primary cosmic ray mass, there is no way to calibrate out \ac{HIM} uncertainties directly.

\section{Developing a Model Suitable for Real Measurements}
\label{sec:FlexibleModel}

In \autoref{sec:Base-training}, networks were trained and predicted on fixed, discrete noise levels.
In real measurements, the noise level will vary and do so even within a single night of observations.
Furthermore, for real data, it may not be practical to first identify the noise levels present in the measurement and then apply a model tailored to that particular noise level, although the idea is interesting. 
Therefore, it is desirable to have a model which can predict \lna{} durably over the full range of expected noise levels. 

It is possible that one of the models described in \autoref{sec:baseperformnacevsNoise} already performs well enough to serve as a usable real-world model.
To identify the best-performing model among all those trained, each model was used to predict \lna{} under all noise conditions.
For each model, the Merit Factors from all noise levels were then averaged to evaluate overall performance.
The results of this study are shown in \autoref{tab:baselinevsgaussian_meritfactor}.
This analysis revealed two promising models:
The 10\,\% baseline / 10\,\% Gaussian noise model (M-B10-G10) performed the best overall; however, this performance was dominated by exceptional predictions of lower-baseline, lower noise showers, but very poor predictions at higher noise levels.
The $75\,\%$ baseline / $20\,\%$ Gaussian noise model (M-B75-G20) conversely excelled in predicting high-baseline high-noise showers, but, surprisingly, performed very poorly in low-noise conditions.
No models from \autoref{sec:baseperformnacevsNoise} performed well across all noise conditions

\begin{table}[t]
    \centering
    \caption{Average Merit Factor achieved by networks trained on different baselines and Gaussian noise levels when evaluated on all other noise levels. The models selected for training the flexible network are highlighted in bold}
    \label{tab:baselinevsgaussian_meritfactor}
    \begin{tabular}{c|cccc}
        \multirow{2}{*}{\shortstack{Baseline\\Height}} & \multicolumn{4}{c}{Gaussian Noise Level} \\
              & 0.025  & 0.050  & 0.100  & 0.200\\ \hline
        0.05  & $1.31 \pm 0.14$  & $1.67 \pm 0.12$  & $1.90 \pm 0.07$ & $1.93 \pm 0.09$\\
        0.10  & $1.07 \pm 0.14$  & $1.73 \pm 0.10$  & \textbf{1.98} $\pm 0.09$  & $1.97 \pm 0.07$\\
        0.25  & $0.82 \pm 0.18$  & $1.02 \pm 0.19$  & $1.87 \pm 0.08$  & $1.94 \pm 0.06$\\
        0.50  & $1.41 \pm 0.12$  & $1.57 \pm 0.10$  & $1.40 \pm 0.18$  & $1.47 \pm 0.18$\\
        0.75  & $0.78 \pm 0.18$  & $0.96 \pm 0.21$  & $1.03 \pm 0.21$  & \textbf{1.76} $\pm 0.05$\\
    \end{tabular}
\end{table}

\subsection{Training a Flexible Network}
\label{sec:GFN}
To develop a model that generalizes across a broad range of noise levels, models M-B10-G10 and M-B75-G20, highlighted in \autoref{tab:baselinevsgaussian_meritfactor}, were both fine-tuned via transfer learning on a newly prepared training dataset, where the noise conditions varied on an event-by-event basis.
In this dataset, for each event, the baseline height was uniformly sampled from $0\,\%$ to $75\,\%$ of peak, while the Gaussian noise was uniformly sampled from $0\,\%$ to $20\,\%$ of peak.
This approach covered all previous noise combinations within these limits, requiring the models to learn how to extract key features without being trained on a fixed noise level.

After training, the average Merit Factor of both models was compared across all noise conditions.
After retraining, the high-noise model, M-B75-G20, achieved consistently high Merit Factors across all noise levels. 
In contrast, even after retraining, the low-noise model, M-B10-G10, continued to perform poorly at the highest noise levels.
The retrained M-B75-G20 also had a significantly higher Merit Factor of the two at all but the lowest noise levels and was therefore selected for further characterization.  
This new flexible model is designated as the Noise Flexible Network (\ac{GFN}) from here on.

\subsection{Performance}
\label{sec:GFNPerformance}
The performance of the \ac{GFN} on its native EPOS-LHC dataset was benchmarked against the fixed noise levels used in \autoref{sec:baseperformnacevsNoise}.
The results of this study are shown in 
\autoref{tab:GFN_EPOS}. 
A key validation of this network is its robust performance over a wide range of noise scenarios, exhibiting a consistent Merit Factor of $\approx 2.2$ across all except the highest noise conditions. 
The level of stability across baseline levels at fixed Gaussian noise, varying by less than $< 1\,\%$ for Gaussian noise levels up to 10\,\% (excepting the most extreme 75\,\% baseline condition), is remarkable.
This suggests that the features driving the prediction for this model are concentrated near the profile peak, where the signal-to-noise ratio is highest regardless of baseline height.
This is encouraging for real measurements, as the peak region is the most reliably observed.
It also raises the question of whether the \ac{GFN} is primarily leveraging information equivalent to the GH shape parameters, which are themselves most constrained near the peak.

\begin{table}[!htb]
    \centering
    \caption{Performance of the \ac{GFN} on different \acp{HIM} and noise conditions (Merit Factor).}
    \label{tab:GFN_comparison}
    \begin{subtable}[b]{0.49\textwidth}
        \centering
        \caption{\ac{GFN} performance: EPOS-LHC}
        \label{tab:GFN_EPOS}
        \begin{tabular}{ccccc}
            \toprule
            \multirow{2}{*}{\shortstack{Baseline\\Height}} & \multicolumn{4}{c}{Gaussian Noise} \\
            \cmidrule(lr){2-5}
             & 2.5\% & 5.0\% & 10\% & 20\% \\
            \midrule
            5\%  & 2.27 & 2.26 & 2.19 & 1.97 \\
            10\% & 2.28 & 2.26 & 2.19 & 2.00 \\
            25\% & 2.28 & 2.26 & 2.17 & 1.97 \\
            50\% & 2.29 & 2.27 & 2.19 & 1.89 \\
            75\% & 2.29 & 2.24 & 2.05 & 1.55 \\
            \bottomrule
        \end{tabular}
    \end{subtable}
    \hfill
    \begin{subtable}[b]{0.49\textwidth}
        \centering
        \caption{\ac{GFN} performance: Sibyll}
        \label{tab:GFN_Sibyll}
        \begin{tabular}{ccccc}
            \toprule
            \multirow{2}{*}{\shortstack{Baseline\\Height}} & \multicolumn{4}{c}{Gaussian Noise} \\
            \cmidrule(lr){2-5}
             & 2.5\% & 5.0\% & 10\% & 20\% \\
            \midrule
            5\%  & 2.07 & 2.06 & 2.01 & 1.84 \\
            10\% & 2.07 & 2.06 & 2.01 & 1.86 \\
            25\% & 2.07 & 2.06 & 2.02 & 1.86 \\
            50\% & 2.08 & 2.06 & 2.01 & 1.82 \\
            75\% & 2.07 & 2.06 & 1.93 & 1.56 \\
            \bottomrule
        \end{tabular}
    \end{subtable}
\end{table}

\subsection{Estimating Hadronic Interaction Model Dependence}
\label{sec:GFNHIM}
To estimate \ac{HIM} model dependence on the performance of the \ac{GFN}, we used the \ac{GFN} to predict on the Sibyll dataset.
\autoref{tab:GFN_Sibyll} shows the results of this cross-prediction for the fixed noise levels. 
A performance degradation, similar to that seen in \autoref{sec:BaseHIM}, is seen, again due to an underprediction of the mass of heavier species.
Despite this, the performance of the \ac{GFN} on Sibyll remains excellent, consistently delivering a Merit Factor of $\approx 2$ across all except the highest noise conditions, again with remarkable stability.
These results further reinforce the potential usability of the \ac{GFN} for real data.
However, again, as there is no way to directly calibrate out these \ac{HIM} uncertainties to set an absolute mass scale for its predictions, the use case for \ac{GFN}-like networks may be strongest for mass ordering.

\subsection{Overall Accuracy of Predictions}\label{sec:NFNAccuracy}

\begin{figure}[!htb]
    \centering
    \includegraphics[width=0.95\textwidth]{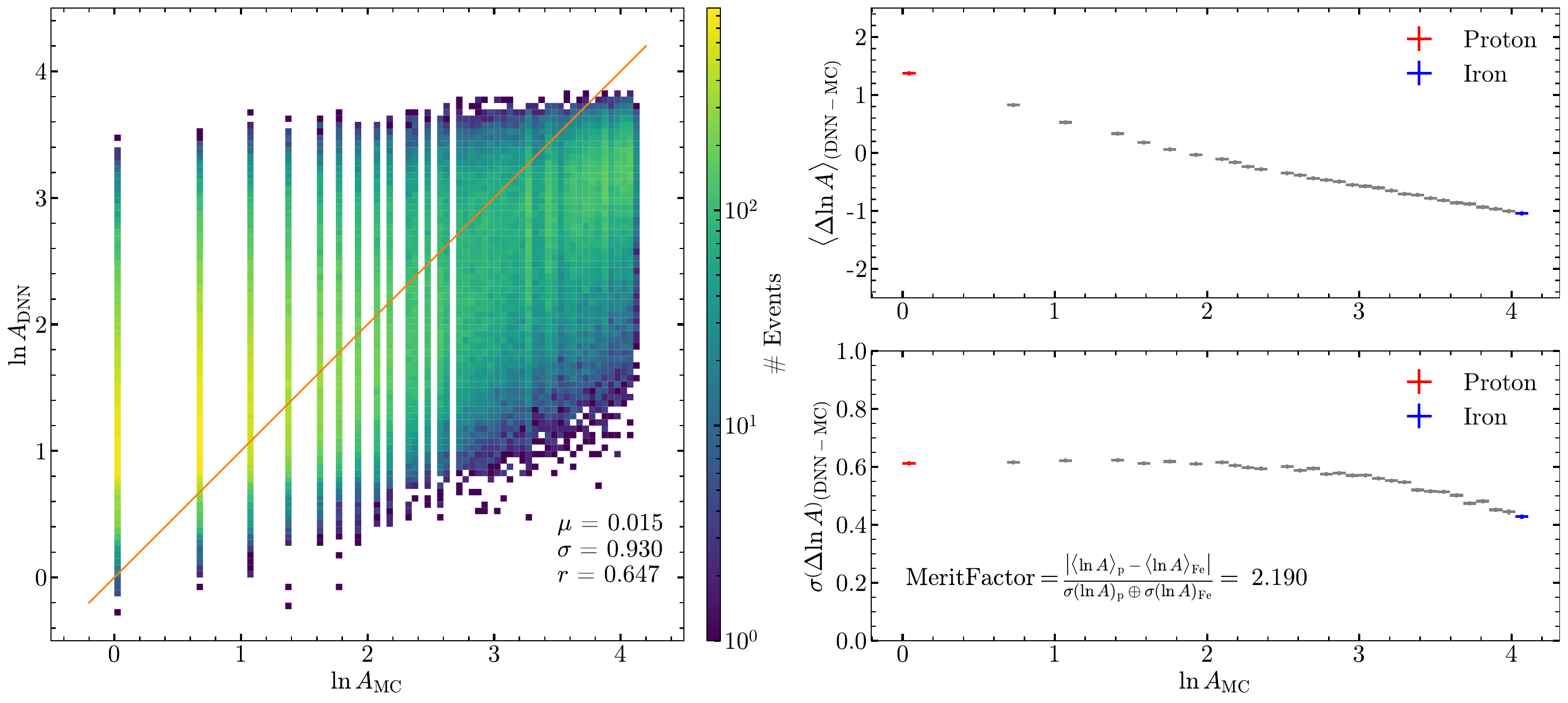}
    \caption{\ac{GFN} Performance on 50\,\% baseline, 10\,\% Gaussian data. Left: Correlation between MC truth \lna{} values and predicted \lna{} values. Right: Bias (top) and resolution (bottom) of the trained network as a function of MC truth \lna{} values. The blue and red data points highlight proton and iron, respectively.}
    \vspace{-2mm}
    \label{fig:GFN_correlation_on_Auger_level_noise}
\end{figure}
To more closely examine the performance of the \ac{GFN}, its predictions of all species should be checked for both separation power and prediction accuracy.  
\autoref{fig:GFN_correlation_on_Auger_level_noise} shows the species-by-species performance of the \ac{GFN} on EPOS-LHC events with the noise level shown in \autoref{fig:auger_noise_profile}.
The left pane of \autoref{fig:GFN_correlation_on_Auger_level_noise} shows the correlation between the \ac{GFN} prediction and truth.
It is clear that the model favors an \lna{} prediction towards the average mass, and therefore it can not be taken 1-to-1 and used as is for further physics analyses. 
The right panes of \autoref{fig:GFN_correlation_on_Auger_level_noise} show the mean bias (top) and resolution (bottom) of the \lna{} predictions.
The mean bias again indicates that the predictions are inaccurate, but well-ordered.

The behavior of achieving low accuracy and decent precision is often seen in ML applications to \ac{UHECR} mass reconstruction, likely due to a combination of factors. 
First, the maximum $A$ value of 61 is close to iron when converting to \lna{}, resulting in edge effects in predictions. 
To address this, future trains are planned on an EPOS-LHC library, which includes elements as high as tellurium (128).
Second, the tendency for ML algorithms to hedge their bets by guessing toward the mean of the training library is often observed.
This behavior is likely exacerbated by our use of \ac{MSE} as the training loss function.
To minimize the prediction collapsing toward the mean, two alternative loss functions are being considered for future models:
\begin{enumerate}
    \item Huber loss: which combines absolute error with \ac{MSE}, giving more weight to outliers and helping to pull the predictions away from mean values.
    \item Negative-Log-Likelihood (NLL): which would predict not just a \lna{} value, but would instead predict mean and variance, which would usefully also provide an uncertainty to the predictions, thereby preventing a hedging toward the mean by allowing the network to include prediction uncertainty.
\end{enumerate}

It is expected that combining the above mitigation techniques will further improve the network's performance.
However, in the interest of timely reporting, we instead apply a linear rescaling of the predicted values to regain accuracy at the cost of precision.
This is done by fitting the mean bias of \lna{} predictions as a function of thrown species, shown in the top right, of \autoref{fig:GFN_correlation_on_Auger_level_noise} with a first-order polynomial of the form 
\begin{align}
    \langle\ln A_\mathrm{DNN}\rangle - \langle\ln A_\mathrm{MC}\rangle &= p_1 \cdot \ln A_\mathrm{MC} + p_0\\
    \Rightarrow \langle\ln A_\mathrm{DNN}\rangle &= (p_1 + 1) \langle\ln A_\mathrm{MC}\rangle  + p_0\\
    \mathrm{where}\quad p_1 &= -0.583\pm0.001\\
    \mathrm{and}\quad p_0 &= 1.215\pm0.003
\end{align}
This relation is then used to calculate a corrected value, $\ln{A_\mathrm{DNN, corr}}$, for each original predicted value $\ln{A_\mathrm{DNN}}$ via
\begin{align}
\ln{A_\mathrm{DNN, corr}} = \frac{\ln{A_\mathrm{DNN}} - p_0}{p_1 + 1}\, .
\end{align}
As the rescaling is linear, it preserves the high Merit Factors of this study while adjusting the means of the predictions for species to the correct values.

\subsection{Performance of the Rescaled \ac{GFN}}\label{sec:NFNPerformance}
\begin{figure}[!htb]
  \centering
  \includegraphics[width=\textwidth]{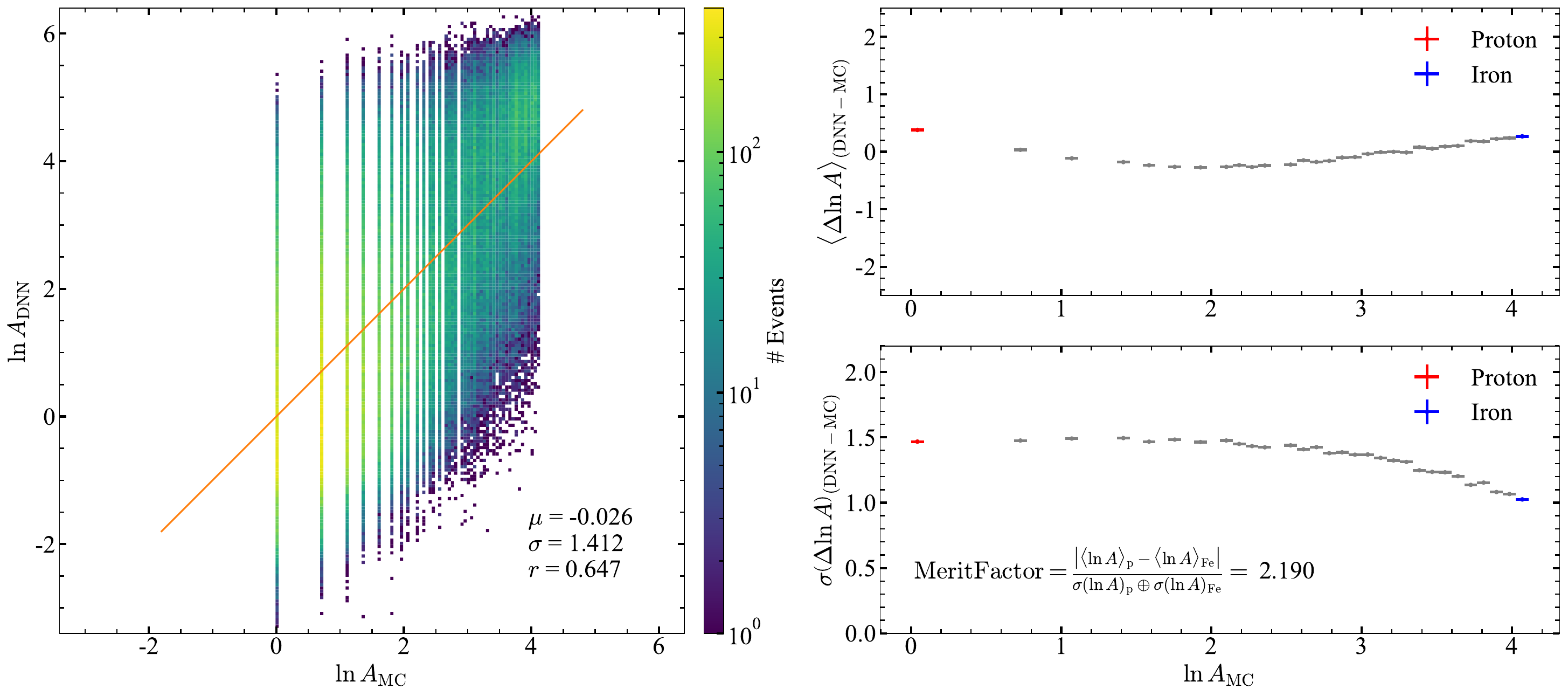}
        \vspace{-2mm}
        \caption{\ac{GFN} Performance on 50\,\% baseline, 10\,\% Gaussian data. Left: Correlation between MC truth \lna{} values and rescaled predicted \lna{} values. Right: Rescaled bias and resolution.}\label{fig:CorrelationScaled}
\end{figure}

\begin{figure}[!htb]
    \centering
    \includegraphics[width=0.45\textwidth]{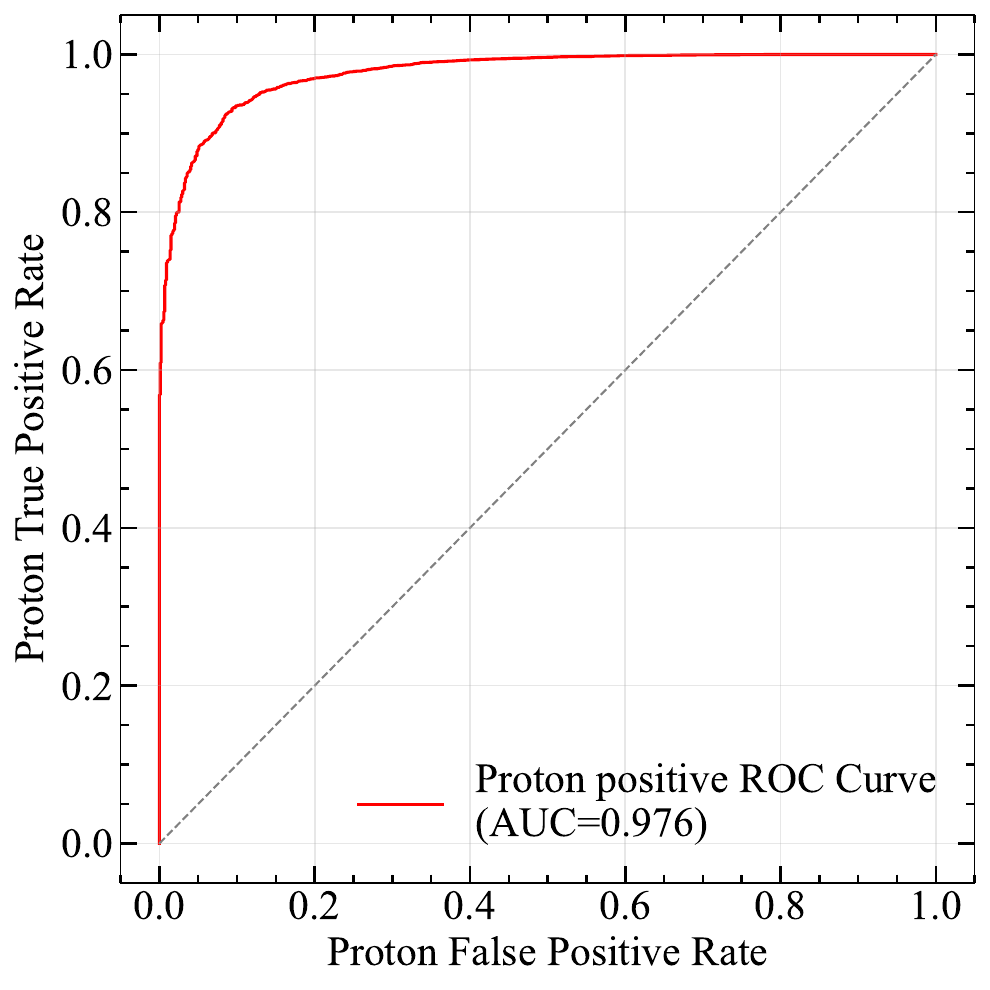}
    \caption{ROC curve assuming proton as the positive. AUC is calculated to be 0.976}
    \label{fig:ROC}
\end{figure}

The results of the above rescaling are shown in \autoref{fig:CorrelationScaled}.
The distributions of the predictions were significantly widened when addressing the prediction biases, which in turn removed the artificially high precision seen in \autoref{fig:GFN_correlation_on_Auger_level_noise}.
\autoref{fig:CorrelationScaled} shows that after this procedure the bias in \lna{} has been mostly removed with 0.4 \lna{} being largest prediction bias seen for any single primary.
With the removal of the bias to the mean, the \lna{} prediction precision can be extracted. The mass of the lightest primaries can be predicted with a resolution of $\sigma(\Delta \ln{A}) \approx 1.5$, while the heavier primaries can be predicted at a higher precision of $\sigma(\Delta \ln{A}) \approx 1$. 
As expected, the Merit factor remains 2.19, indicating the mass separation power has been preserved in the rescaling.
To provide a threshold-independent gauge of the proton--iron separation implied by the predicted \lna{} distributions, we also construct the proton-positive ROC curve from the \ac{GFN} predictions.
The resulting curve is shown in \autoref{fig:ROC}, with $\mathrm{AUC} = 0.976$, consistent with the Merit Factor of 2.19.

\subsubsection*{Evaluating \ac{GFN} Performance on Intermediate Masses}

\begin{wrapfigure}{r}{0.4\textwidth}
    \captionsetup{type=table}
    \centering
    \vspace{-4mm}
    \caption{Figures Of Merit from intermediate mass pairings.}
    \begin{tabular}{c|cccc}
        $_{A1} \backslash ^{A2}$ & p  & $^4$He  & $^{14}$N  & $^{56}$Fe \\ \hline
        p  & 0    & 0.39 & 1.04 & 2.19 \\
        $^4$He & 0.39 & 0    & 0.63 & 1.71 \\
        $^{14}$N  & 1.04 & 0.63 & 0    & 1.04 \\
        $^{56}$Fe & 2.19 & 1.71 & 1.04 & 0    \\
    \end{tabular}
    \label{tab:MFTable}
    \vspace{-3mm}
\end{wrapfigure}
Though the Merit Factor is frequently used to compare the mass separation power between methods, it is an incomplete measure.
First, the measurements of the mass composition of cosmic rays at the Pierre Auger Observatory do not suggest there will be many opportunities to distinguish a proton subsample in a flux that is, in the majority, made of iron primaries~\cite{PierreAuger:2025rdo}.
Instead, it is more likely that heavy and light components between these two extremes will need to be distinguished from each other.
To this end, the $FOM$s for the pairings of proton ($A=1)$, helium ($A=4)$, nitrogen ($A=14$), and iron ($A=56$), the four commonly used mass groups, are calculated using \autoref{eq:mf} and are shown in \autoref{tab:MFTable}.

By examining \autoref{tab:MFTable}, it is clear that it would be much easier to remove high mass primaries from a lighter sample than it would be to remove lower mass primaries from a heavy sample. 
For example, separating proton and nitrogen ($\Delta \ln{A} = 2.63$) can be done with $FOM(1,14)=1.04$, which is the same value obtained when nitrogen and iron ($\Delta \ln{A} = 1.39$) are instead used.
Luckily, this is typically not an issue for most analyses, as light samples of high purity are often the goal rather than high-purity heavy samples.

\begin{figure}[!b]
    \centering
    \begin{subfigure}[t]{0.52\textwidth}
        \centering
    \includegraphics[height=60mm]{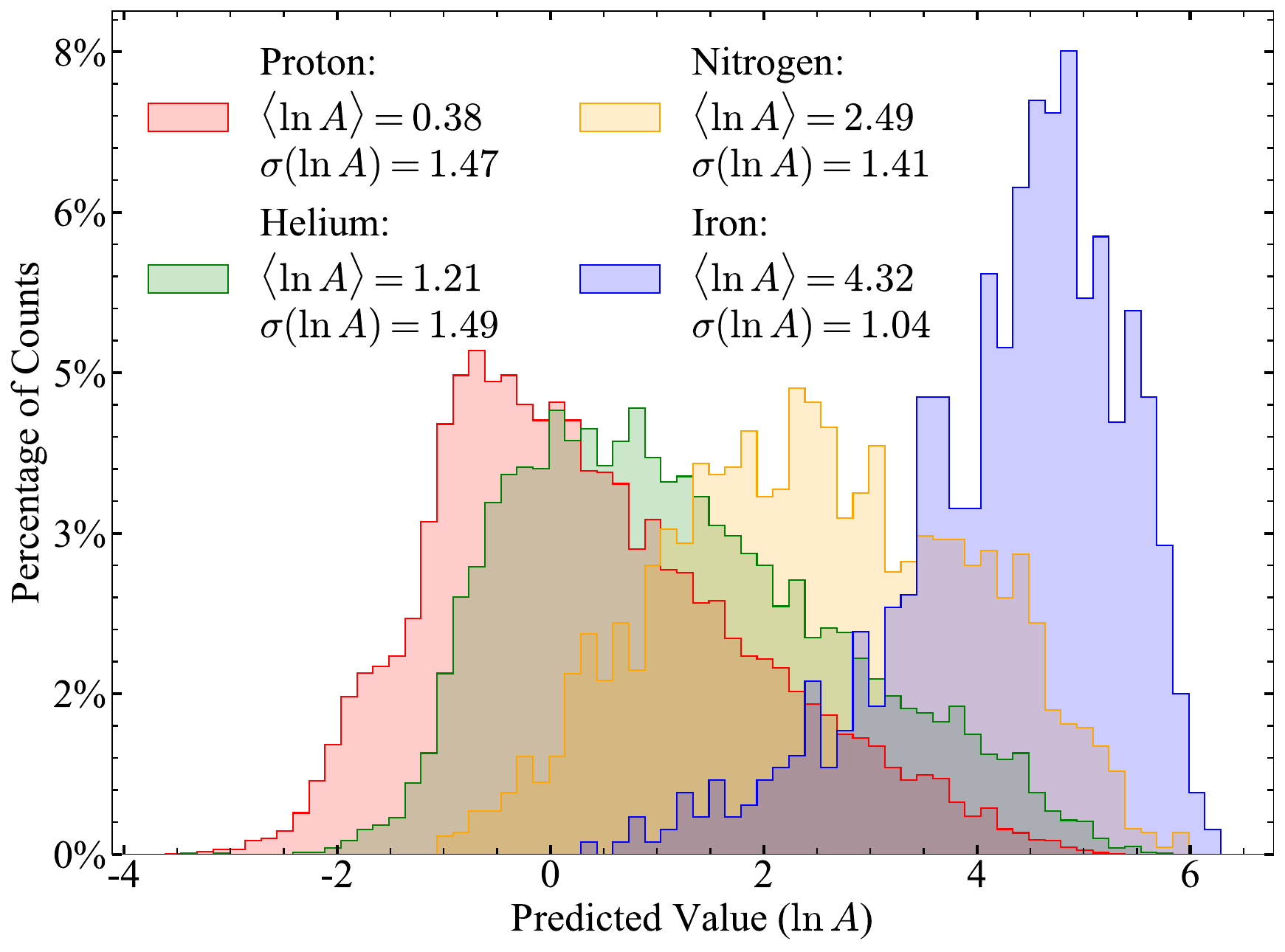}
        \caption{Predicted \lna{} for p, He, N, and Fe.}
        \label{fig:PredictionsScaled}
    \end{subfigure}
    \hfill
    \begin{subfigure}[t]{0.45\textwidth}
        \centering
        \includegraphics[height=61mm]{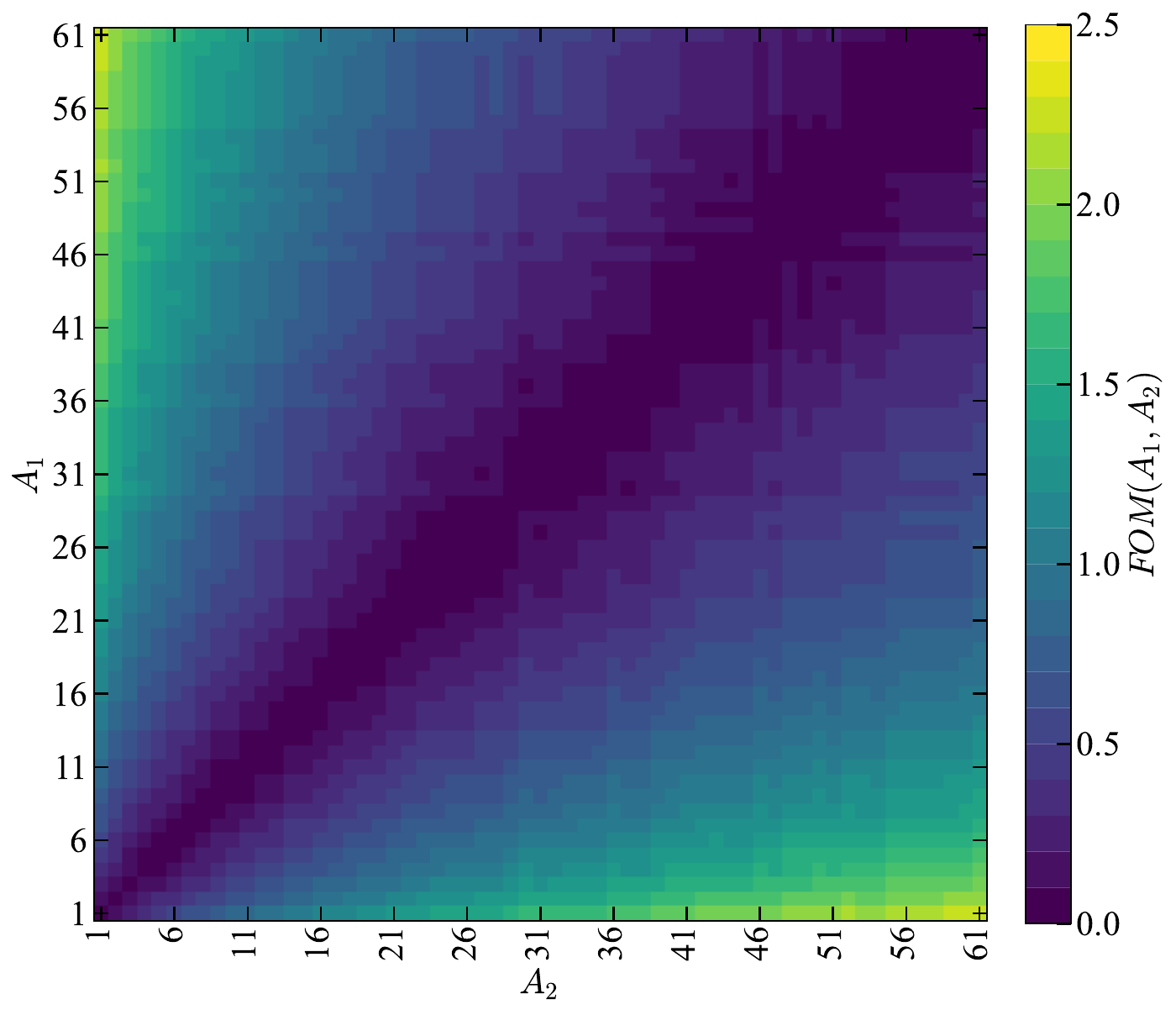}
        \caption{$FOM$ for all parings.}
        \label{fig:confusion_matrix}
    \end{subfigure}
    \caption{Performance of the \ac{GFN} on intermediate mass primaries. a) shows the predictions of the \ac{GFN} on four representative primaries, p, He, N, Fe. b) shows the $FOM$ for all available pairings of $A$ between $A=1$ and $A=61$.}
    \label{fig:ROC+Matrix}
\end{figure}

This general behavior of harder-to-distinguish light primaries is not unexpected, as lower-mass primaries are subject to higher shower-to-shower fluctuations, which would affect any mass-sensitive shower observable.
This interpretation is somewhat borne out by examining \autoref{fig:PredictionsScaled}, which shows the distributions of predicted \lna{} for each of the four above-listed primaries.
First, as expected, protons and helium display longer tails to higher \lna{} values.
The higher shower-to-shower fluctuations are surely causing much of these tails; however, surprisingly, iron also displays a tail extending into lighter \lna{} values.
This unexpected iron tail suggests that, in addition to shower-to-shower fluctuations, the network is also still inclined to predict the mean and that primaries near the edge of the training sample are most severely affected by this effect. 
These distributions additionally illustrate that the Figure Of Merit is an imperfect descriptor of mass-sensitive parameters, as their distributions tend to be non-normal and asymmetric.
However, since the bulk of published research utilizes the $FOM$, it remains a useful metric for comparing the ML method described here to other methods and approaches.
Last, for completeness, the Figure of Merit for all possible combinations of primaries with mass numbers between 1 and 61 is shown in \autoref{fig:confusion_matrix}.

\subsubsection*{Robustness of the Merit Factor as a Performance Metric}

To further characterize the Merit Factor as a performance metric in this context, we also
examined whether the Merit Factor values reported throughout this paper could be susceptible
to the variable-change inflation identified in \autoref{sec:baseperformance}.
To test this, we trained the linear discriminant, XGBoost, and the CNN on a bimodal
proton/iron-only training set and compared the resulting Merit Factor and AUC to those
of the corresponding All-$A$ trained models.
The results are shown in \autoref{tab:BimodalVsContinuous}.
 
\begin{table}[ht]
\centering
\caption{Merit Factor and AUC on noise-free profiles with bins prior to
$X_{\rm first}$ blinded, comparing All-$A$ to bimodal p/Fe-only training.}
\label{tab:BimodalVsContinuous}
\renewcommand{\arraystretch}{1.3}
\begin{tabular}{l l cc}
\hline\hline
\multirow{2}{*}{Model} & \multirow{2}{*}{Metric}
  & \multicolumn{2}{c}{Noise-free} \\
\cline{3-4}
 & & All $A$ & p/Fe only \\
\hline
\multirow{2}{*}{Linear Discriminant}
  & Merit Factor & 1.69   & 1.75   \\
  & AUC & 0.983  & 0.983  \\
\hline
\multirow{2}{*}{XGBoost (4 params)}
  & Merit Factor & 2.83   & 3.51   \\
  & AUC & 0.992  & 0.992  \\
\hline
\multirow{2}{*}{CNN}
  & Merit Factor & 6.83   & 12.80  \\
  & AUC & 0.9999 & 0.9995 \\
\hline\hline
\end{tabular}
\end{table}

Merit Factor inflation under bimodal training scales with model expressiveness: the linear discriminant shows negligible inflation ($+4\%$), XGBoost shows moderate inflation ($+24\%$), and the CNN shows substantial inflation ($+87\%$).
Crucially, the AUC does not follow: for the CNN it is essentially unchanged and marginally decreases ($0.9999 \to 0.9995$) under bimodal training, confirming that the Merit Factor increase reflects distortion of the intermediate-mass variable space rather than a genuine improvement in discriminating power.
In our main results, all models are trained on the full All-$A$ distribution, which suppresses this mechanism entirely as any mapping that distorted intermediate-mass predictions would be penalized by the MSE loss on those species at every training step.
This is further evidenced by the fact that Merit Factor and AUC rank all three models identically in the All-$A$ column (Linear Discriminant $<$ XGBoost $<$ CNN), consistently across noise conditions, confirming that the two metrics are in agreement
and that the Merit Factor values reported throughout this paper reflect genuine differences in discriminating power.

\subsection{Comparison to Gaisser-Hillas Shape Parameter Models}\label{sec:GHComparison}

To place the \ac{GFN} performance in direct comparison with GH-parameter-based approaches on the same dataset, we implemented a Gaisser-Hillas fit following the procedure of~\cite{PierreAuger:2018gfc} and extracted \xmax{}, $E_{\rm cal}$, $L$, and $R$ from profiles at the typical noise level.
A straight fit to the data is poorly constrained at high baseline levels, as the wings of the profile fall below the noise floor, leading to large biases in $L$ and $R$ in particular.
To address this, we also implemented an improved \emph{peak-window} fit that restricts the fitting range to the $\pm 25$ bins surrounding the profile peak, where the signal reliably exceeds the 50\,\% baseline.
The peak-window fit substantially reduces the bias and resolution of \xmax{} and $E_{\rm cal}$, though a large bias in $R$ persists due to the restricted fitting range truncating the profile wings where $R$ is most sensitive.
The bias and resolution of the extracted parameters under both procedures are compared in \autoref{tab:GHBiasResolution}.

\begin{table}[ht]
\centering
\caption{Bias and resolution of GH parameters extracted from profiles at the typical
noise level (50\,\% baseline, 10\,\% Gaussian), for the standard fit and the
peak-window fit restricted to $\pm 25$ bins around the profile peak.}
\label{tab:GHBiasResolution}
\renewcommand{\arraystretch}{1.3}
\begin{tabular}{l l cc}
\hline\hline
Fit method & Parameter & Bias (\%) & Resolution (\%) \\
\hline
\multirow{4}{*}{Standard fit}
  & \xmax{}                  & $+2.08$   & $19.0$  \\
  & $\log_{10}E_{\rm cal}$   & $+2.13$   & $0.944$ \\
  & $L$                      & $+162$    & $21.0$  \\
  & $R$                      & $+80.7$   & $21.4$  \\
\hline
\multirow{4}{*}{Peak-window fit}
  & \xmax{}                  & $-0.376$  & $1.78$  \\
  & $\log_{10}E_{\rm cal}$   & $+0.0817$ & $0.322$ \\
  & $L$                      & $+3.48$   & $11.2$  \\
  & $R$                      & $+39.1$   & $51.5$  \\
\hline\hline
\end{tabular}
\end{table}

A linear discriminant and an XGBoost model were then trained to predict $\ln A$ from the peak-window extracted parameters, representing the strongest available GH-parameter baseline.
Their performance relative to the \ac{GFN} is shown in \autoref{tab:RealisticNoiseComparison}.

\begin{table}[ht]
\centering
\caption{Proton--iron Merit Factor and AUC on typical noise data (50\,\% baseline, 10\,\%
Gaussian), comparing models trained on peak-window extracted GH parameters (\xmax{}, $E_{\rm cal}$, $L$, and $R$) to the \ac{GFN}.
GH-parameter models use the peak-window fit from \autoref{tab:GHBiasResolution}.}
\label{tab:RealisticNoiseComparison}
\renewcommand{\arraystretch}{1.3}
\begin{tabular}{lcc}
\hline\hline
Model & Merit Factor & AUC \\
\hline
Linear Discriminant (4 par.) & 1.49 & 0.973 \\
XGBoost (4 par.)             & 2.14 & 0.975 \\
\ac{GFN} (full profile)                                        & 2.19 & 0.976 \\
\hline\hline
\end{tabular}
\end{table}

With the improved peak-window extraction providing near-unbiased \xmax{} and $E_{\rm cal}$, the \ac{GFN} continues to outperform the best GH-parameter model in both Merit Factor and AUC.
Notably, however, the XGBoost model trained on only four extracted GH parameters already achieves a Merit Factor of 2.14, substantially exceeding both the realistic-resolution (\xmax{}, $R$, $L$) benchmark of 1.51~\cite{Flaggs:privcomm} and the $N_\mu$-based Merit Factor of 1.8~\cite{Flaggs:2023exc}.
This demonstrates that applying modern ML methods to already-reconstructed profile-shape parameters offers a practical, immediately deployable path to improved mass sensitivity in existing observatories.
The closeness of the XGBoost and \ac{GFN} Merit Factors and AUC at typical noise raises the question of whether the \ac{GFN} is primarily leveraging information equivalent to the GH parameters under realistic conditions.
The ablation study in \autoref{sec:massinfo} demonstrates that the profile does contain composition-sensitive structure beyond the GH parameterization on noise-free data; however, whether this additional information remains accessible at typical noise is not yet established.
Regardless, the \ac{GFN} carries a practical advantage: it operates directly on the raw profile without requiring a separate fitting step whose quality degrades with noise, and maintains stable performance across the full range of tested noise conditions without any adjustment (see ~\autoref{tab:GFN_EPOS}).
This robustness of the CNN would be difficult to match with a method dependent on GH parameter extraction under highly variable observing conditions.

\section{Discussion}\label{sec:discussion}
Machine learning methods are perhaps a bit en vogue in astroparticle physics currently and can be seen as a method in search of a problem.
However, in the case of air-shower development profiles, current practice typically relies on \xmax{} alone or in combination with energy, leaving most of the composition-sensitive information in the profile unused. 
This presents a clear opportunity for ML methods to excel.
Indeed, this work demonstrates that deep-learning methods hold substantial promise for significantly enhancing particle identification using air-shower development profile data.
Using the Merit Factor and AUC as gauges, the CNN  developed in this work outperforms both a linear discriminant and an  XGBoost model trained on GH parameters extracted from the same data (see ~\autoref{tab:RealisticNoiseComparison}), while the XGBoost result itself already substantially exceeds published benchmarks for realistic-resolution combinations of \xmax{}, $R$, and $L$ (1.51~\cite{Flaggs:privcomm}) and for $N_\mu$ alone (1.8~\cite{Flaggs:2023exc}).
Furthermore, the CNN maintains this performance on profiles noised far beyond what is commonly accepted in composition analyses at modern observatories.
Moreover, on noise-free data the CNN achieves a Merit Factor of 6.83, far exceeding the 2.83 obtained by XGBoost on the same events' GH parameters (\autoref{tab:NoiseFreeComparison}), providing a first indication that the full profile carries composition-sensitive information beyond what the GH parameterization captures.
These observations show that Machine-learning algorithms are well-suited to combining all information available in a profile at once to predict the mass of individual \ac{UHECR} primaries. 

\subsection{Mass Information Beyond the Standard Profile-Shape Observables}
\label{sec:massinfo}
To quantify how much more information may be present and usable in shower profiles, we carried out an ablation study in which raw profile bins were progressively added to the GH parameters as inputs to an XGBoost model, on noise-free data. The additional bins were sampled evenly from the full 700-bin profile, first at 10\% (every tenth bin) and then at 100\% (all bins). 
The results are shown in \autoref{tab:Ablation}.

\begin{table}[ht]
\centering
\caption{Merit Factor and AUC on noise-free data as a function of the inputs
provided to XGBoost, progressing from the four extracted GH parameters (\xmax{}, $E_{\rm cal}$, $L$, and $R$) alone to
those parameters supplemented with increasing numbers of raw profile bins.
The CNN trained on the full profile (after $X_{\rm first}$) from \autoref{sec:noisefree} is shown for reference.}
\label{tab:Ablation}
\renewcommand{\arraystretch}{1.3}
\begin{tabular}{lcc}
\hline\hline
XGBoost input & Merit Factor & AUC \\
\hline
4 GH params only                & 2.83  & 0.992  \\
4 GH params + 10\% profile  & 3.79  & 0.999  \\
4 GH params + full profile & 5.58  & 0.9999 \\
CNN (full profile)              & 6.83  & 0.9999 \\
\hline\hline
\end{tabular}
\end{table}

Each step in \autoref{tab:Ablation} holds the GH parameters fixed and adds only raw profile information.
This means any performance gain must arise from composition-sensitive structure in the profile that the GH functional form discards rather than from improvements in fit quality.
The performance of XGBoost increases monotonically and substantially as the profile data is added, confirming that the GH parameterization is informationally lossy with respect to composition.
Furthermore, even when XGBoost has access to the full profile alongside the GH parameters, it is still outperformed by the CNN (Merit Factor 5.58 vs.\ 6.83), demonstrating that the CNN's ability to learn local and hierarchical features from sequential data is better matched to this task than gradient-boosted trees.

In addition to the direct application of ML methods to the full recorded profile,
future work should also aim to identify which aspects of the profile drive this
additional separation power and whether they can be summarized into robust observables
for incorporation into more traditional composition analyses.

\subsection{The Suitability of our Architecture for Full Profile Mass Predictions}
This study evaluated several approaches to predicting $\ln A$ from shower profile data, spanning a range of model complexity.
At the simplest end, a linear discriminant trained on four extracted GH parameters provides a straightforward baseline; XGBoost on the same 
parameters offers a strong non-linear alternative that is competitive with the CNN at typical noise levels (\autoref{tab:RealisticNoiseComparison}).
Among the deep-learning architectures tested (LSTM, Transformers, and 1D-CNN), the 1D-CNN demonstrated the highest performance plateau and the fastest training time for this task.
We do not assert that the developed and tested network represents the best choice for this particular application; however, we found the lightness and flexibility of a CNN to be particularly useful for gauging the general potential of the method.

To evaluate the role of model complexity, we trained a shallower CNN by removing block 2 and block 4 from the original architecture shown in \autoref{tab:CNN_architecture}.
The reduced network achieved performance nearing that of the original model on the 50\,\%/5\,\% baseline/Gaussian benchmark, with a Merit Factor of 2.22 (versus 2.39) and best validation losses of 0.8283 (versus 0.8199).
This suggests that the dominant mass-sensitive features can be extracted with a lighter-weight architecture.
Differences in training behavior were, however, observed.
The reduced network became unstable after approximately 1600 epochs and began to skew toward mean prediction, whereas the original architecture remained stable for roughly 2500 epochs.
In addition, the reduced model reached its minimum validation loss quicker at epoch 770, while the original model continued to improve until epoch 2368.
Although stochastic training effects might influence these values, the deeper architecture appears more stable, able to capture more fine-grained details, and more capable of deeper refinement.

While training, we found that to build a network that could robustly predict in high noise levels, the noise level had to be gradually increased, as networks failed to converge when we started at high noise levels. 
This was achieved by initially training on low noise levels and then transfer learning to increasingly higher noise levels.
We then found that to produce a network that could stably predict over a large range of noise levels, we had to take a model trained on very high noise levels and then train it on variable noise.
We expect further improvements can be made to both the network architecture and training approach, which would result in better accuracy and separation power.
However, this serves to underscore the central finding that ML methods can and should be applied to air-fluorescence data to extract detailed composition information on primary cosmic rays.

The strong performance of XGBoost on extracted GH parameters at typical noise (\autoref{tab:RealisticNoiseComparison}) demonstrates that the CNN's advantage at realistic noise levels is modest.
However, the CNN retains clear architectural advantages for this task: it requires no separate parameter extraction step, it achieves substantially higher performance on clean data (\autoref{tab:NoiseFreeComparison}), and the \ac{GFN} maintains stable performance across a wide range of noise conditions without adjustment (\autoref{tab:GFN_comparison}).
The practical implications of these complementary strengths for current and future observatories are discussed in \autoref{sec:outlook}.

\subsection{Potential Applications and Impact}
\label{sec:outlook}
The application of machine learning techniques to the reconstruction of \ac{UHECR} longitudinal profiles offers two complementary avenues for advancing composition studies. 
First, gradient-boosted decision tree models such as XGBoost, applied to already-reconstructed GH parameters, can take immediate advantage of parameter values that already exist in current observatory databases, requiring no reprocessing of raw data.
As shown in \autoref{tab:RealisticNoiseComparison}, this approach already exceeds published benchmarks, offering an immediately deployable improvement in mass sensitivity.
Second, applying a CNN directly to the full profile provides the highest demonstrated sensitivity and avoids dependence on a separate fitting step, making it the stronger long-term approach, particularly under variable or challenging noise conditions. 
Both paths should be pursued using the fluorescence telescope data of Pierre Auger Observatory~\cite{PierreAuger:2009esk} and Telescope Array Observatory~\cite{TelescopeArray:2008toq}.
To accomplish this, CONEX simulation libraries similar to those described in \autoref{sec:dataset} should be produced using updated models and extended to higher masses, then processed through each observatory's detector simulation framework 
to produce realistic training datasets.
Performance should be benchmarked using these simulation sets, and the networks should then be used to extract \lna{} predictions on real data.

Unfortunately, both approaches inherit dependence on hadronic interaction models.
For the CNN, this dependence has been partially characterized: the \ac{GFN} trained on EPOS-LHC retains strong separation power when cross-predicting on Sibyll (\autoref{tab:GFN_Sibyll}), though predictions for heavier species shift systematically to lighter values.
For the XGBoost approach, the mapping from GH parameters to $\ln A$ learned during training is itself HIM-dependent, as the values of \xmax{}, $L$, and $R$ for a given primary shift between models.
While the GH parameterization itself fits the average longitudinal profile to within 1\,\% regardless of HIM choice~\cite{PierreAuger:2018gfc}, the trained relationship between these parameters and mass has not been cross-validated between models, making this an important open question for deployment on real data.
Because $\ln A$ is directly reconstructed in both approaches, calibrating out HIM-dependent biases would require identifying reference samples of known composition in the data.
While proton-enriched samples may be identifiable at lower energies, it is unlikely that such a sample can be constructed across the full energy range, and there is no well identified heavy primary that could stand in at high energies.
A careful study of HIM-dependent biases using post-detector simulations is therefore needed for both approaches before deployment on real data. 
Despite these challenges, both ML-aided approaches hold promise to provide significant light/heavy separation power.

\paragraph{Outlook for next-generation space-based observatories}
Proposed space-based missions, such as \ac{POEMMA} and \ac{MEUSO}, as well as future balloon-borne detectors like \ac{PBR}, stand to gain substantially from machine learning-driven mass reconstruction. 
The observatories measure \ac{EAS} via air-fluorescence from a vantage point above the Earth's atmosphere, enabling a large instantaneous aperture.
However, in contrast to the ground-based observatories, which can employ hybrid methods to reconstruct the mass composition of primaries, space-based observatories can only extract mass data using the shower profile.
Typically, this meant relying on extracted summary parameters such as \xmax{}, $R$, and/or $L$ alone.
This in turn meant that the maximum achievable Merit Factor would be the 1.6 predicted in~\cite{Flaggs:2023exc}, a level far below the minimum Merit Factor of 2.0 needed for backtracking studies and charged particle astronomy in~\cite{Coleman:2022abf}.
While the XGBoost approach formally exceeds this threshold on ground-based data (Merit Factor 2.14, \autoref{tab:RealisticNoiseComparison}), reliable extraction of GH parameters from space-based profiles will be considerably more challenging due to lower signal-to-noise ratios and more limited profile coverage, as illustrated by the sensitivity of the GH fit to noise conditions shown in \autoref{tab:GHBiasResolution}.
Conversely, the \ac{GFN}'s ability to operate directly on the raw profile without a separate fitting step and its robustness against a very wide range of noise levels makes it particularly well-suited to this use case.
From this work, it is clear that networks like the \ac{GFN} may be able to help make space-based detectors competitive not only from an exposure point of view, but also in terms of composition measurement.
This possibility is further enhanced due to the distinct possibility that HIMs will significantly improve by the time the next generation of detectors is built or flown.
This means the issues with the difficult-to-calibrate-out HIM uncertainties may be entirely resolved by the time space-based data is ready for a full profile DNN reconstruction, further enhancing the possible mass resolution of a space-based detector. 

\paragraph{Outlook for Next-Generation Ground-based Observatories}
Proposed next-genera-tion ground-based observatories,  such as \ac {GCOS}~\cite{Fujii:2025ytc}, aim to cover vast areas (up to 60,000 km$^2$) and are targeting very high composition sensitivity to reach the desired exposure goals.
This work, if it indeed translates to real event reconstruction, should aid considerably in the pursuit of these goals.
Currently, even the high Merit Factors achieved here or the ideal combination of \xmax{} and \Nmu{} can not alone achieve the targeted $\sigma(\Delta \ln A) = 1.0$ resolution~\cite{Fujii:2025ytc}.
However, combining a DNN predicting on the full shower profile with the independent reconstruction of \Nmu{} in a hybrid observation may allow for this goal to be met.
Moreover, the ground-based case is particularly favorable for the XGBoost approach, as GH parameter reconstruction is already routine in existing observatory pipelines and incorporating additional observables such as $N_\mu$ as input features is straightforward.
Even the XGBoost, when combined with an independent $N_\mu$ measurement, may approach the GCOS target, while the \ac{GFN} could possible push sensitivities further still.
Furthermore, as stated above, HIM uncertainties may be resolved by the time the next generation of detectors is taking data, meaning the performance and accuracy degradation seen when moving between models may be entirely avoided.
Likewise, the capabilities of ML analysis approaches will continue to increase in power during the design, development, and commissioning of a next-generation ground-based \ac{UHECR} observatory, meaning even better predictive performance than what is described here can be expected in the future.

\acknowledgments
We would like to acknowledge the funding provided by the 23/24 and 24/25 Colorado School of Mines Undergraduate Research Fellowships (MURF) for students Z. Wang, N. Woo. and C. Smith.
We would like to acknowledge the support of NSF award PHY2310111, which supported the time of Dr. S. Mayotte.
We would like to acknowledge the support of the NASA awards 80NSSC22K1488/80NSSC24K1780, which supported Dr. E. Mayotte and J. Burton during the study.
We would also like to acknowledge the Colorado School of Mines startup funds, which supported the time of N. San Martin and Dr. E. Mayotte, as well as the computation resources used in this study.
Finally, we would like to acknowledge those who provided feedback on the pre-publication manuscript:
Michael Unger for his insight into the possible sources of the extreme merit factor achieved by the noise-free model.
Benjamin Flaggs for his comments, questions, and clarifications, but especially for furnishing supplemental information to~\cite{Flaggs:2023exc}.
We also thank the anonymous referee, whose feedback greatly strengthened the paper.

\bibliographystyle{elsarticle-num}
\bibliography{References}

\end{document}